# Advances and opportunities for automated robotic preparation of 2D materials and fabrication of 2D heterostructures

Shiva Davari[1,2], Dinh Loc Duong[1,3], Tim Faltermeier[1,4], Josue A. Goss[1,2], Amirhossein Hasani[1,4], Dharmraj Kotekar-Patil[1,2], Jane Peabody[1,5], Samuel Wyss[1,4], Hugh O. H. Churchill[1,2,*], Nicholas J. Borys[1,4,5,†]

[1]MonArk NSF Quantum Foundry

[2]Department of Physics, University of Arkansas, Fayetteville, AR

[3]Department of Physics and Frontier Institute for Research in Sensor Technologies, University of Maine, Orono, ME

[4]Department of Physics, Montana State University, Bozeman, MT

[5]Department of Physics, University of Utah, Salt Lake City, UT

**Abstract:** The mechanical exfoliation, transfer, and stacking of 2D atomic sheets from layered van der Waals crystals synergize to enable atomic layer-by-atomic layer engineering of 2D heterostructures with precisely tailored physical properties that give rise to new exotic phenomena and states of matter. With the huge variety of layered van der Waals materials available, there is a limitless number of ways to couple disparate 2D semiconducting, insulating, magnetic, metallic, topological, etc. systems with one another. Experimental exploration of this vast space starts with the fabrication of high-quality 2D heterostructures, which is commonly performed manually, relying on humans to execute delicate operations. Many significant scientific advancements have been achieved in this manner, revealing immense potential for further discovery and innovation of increasingly sophisticated 2D heterostructures. However, soon, the complexity of the 2D heterostructures that define the state-of-the-art for scientific discovery will exceed the capabilities of manual fabrication. Therefore, the demand for robotic instruments for preparing 2D materials and fabricating complex 2D heterostructures with greater quality, at higher rates, and with better reproducibility is rapidly increasing. This review covers the most recent scientific, instrumentation, and processing advances rising to this challenge. Robotic instruments for mechanical exfoliation, optical metrology of 2D crystallites, stacking, as well as key advancements in supporting technologies such as organic-free stamps, vacuum-compatible processing tools, and artificial intelligence are covered. Looking forward, a new generation of artificial intelligence-driven, automated advanced manufacturing tools is anticipated to emerge from these current advancements. These new tools will bridge the current state-of-the-art of 2D heterostructure science to new scientific frontiers defined by precision fabrication of high-quality, complex, many-layer 2D heterostructure systems.

* hchurch@uark.edu
† nicholas.borys@utah.edu

## 1. Introduction

Advances in the formation of 2D material heterostructures—artificial assemblies of similar and dissimilar layered materials informally referred to as “stacks”—have continually reinvigorated the field of 2D materials and their applications for 15 years since graphene was first laminated over hexagonal boron nitride (hBN)[1]. hBN proved to be an ideal, atomically flat substrate for graphene, dramatically improving its electronic transport properties. Another step change in graphene quality came when it was discovered that fully encapsulated graphene in hBN could be contacted well along the one-dimensional edge exposed by etching through the stack[2]. Next, by encapsulating inside an argon- or nitrogen-filled glovebox, air-sensitive monolayer 2D ferromagnets[3,4], semiconductors[5], and superconductors[6] could be studied for the first time.  Graphene device quality improved again when 3D gate metals were replaced with graphite gates[7].

The menu of physical phenomena that could be explored expanded enormously by moving beyond inert substrates and combining multiple functional materials within a heterostructure. Examples include superconductivity in magic-angle twisted graphene[8]; proximity-induced spin-orbit coupling[9], ferromagnetism[10], and superconductivity[11]; and emergent ferroelectricity in homobilayers[12], naming just a few among many. Twisted heterostructures have become a particularly fruitful platform to explore exotic physical phenomena, including superconductivity[13], magnetism[14], Wigner crystals[15,16], and integer/fractional quantum anomalous Hall effects[17-24].

In all cases, these discoveries were enabled by advances in heterostructure fabrication technologies to create ever cleaner and more perfect interfaces with more precise orientations of the layers. Given the painstaking nature of manual heterostructure assembly, researchers in the field have recently emphasized the importance of using automation, machine learning and artificial intelligence (AI), and robotics to accelerate progress. In this review, we discuss some recent developments in this technology, with an emphasis on methods to accelerate, automate, and continually improve the stacking process. Our primary focus is stacking of flakes mechanically exfoliated from bulk crystals, although in many cases, similar techniques can be adapted to manipulate large-area 2D crystals grown by various techniques[25]. We begin with an overview of heterostructure assembly in Section II. Section III discusses automated exfoliation to produce 2D material flakes. Section IV discusses methods for automated flake detection and metrology using machine learning methods, followed by a review of robotic stacking in Section V. Section VI covers assembly of heterostructures in glovebox environments, and ultrahigh vacuum (UHV) stacking is reviewed in Section VII. Section VIII highlights the use of inorganic stamps rather than polymers for stacking, and we conclude with an outlook for this field. Sections IX and X conclude with our perspective on progress towards full automation and outlook for future efforts to enhance the robotic preparation of high-quality 2D materials, heterostructures, and devices.

## 2. Overview of the fabrication of 2D heterostructures

Fabricating high-quality 2D heterostructures requires painstaking attention to detail to achieve the well-defined interlayer twists, strong interlayer couplings, and minimal disorder that are needed to observe the exotic phenomena and new states of matter that they host. Along with precise interlayer alignment, high-quality van der Waals heterostructures rely on atomically clean and residue-free interfaces for optimal performance. Superficially, the assembly process appears simple: mechanically exfoliate (or directly grow) the desired crystallites for the heterostructure; record their positions, shapes, and crystallographic orientations; use computer-aided design (CAD) software to position and orient each flake to each other; and then, using a stamp with tunable adhesion (usually elastomer-based) combined with micromanipulation tools, pick-up each layer and stack them onto a target substrate (Figure 1).

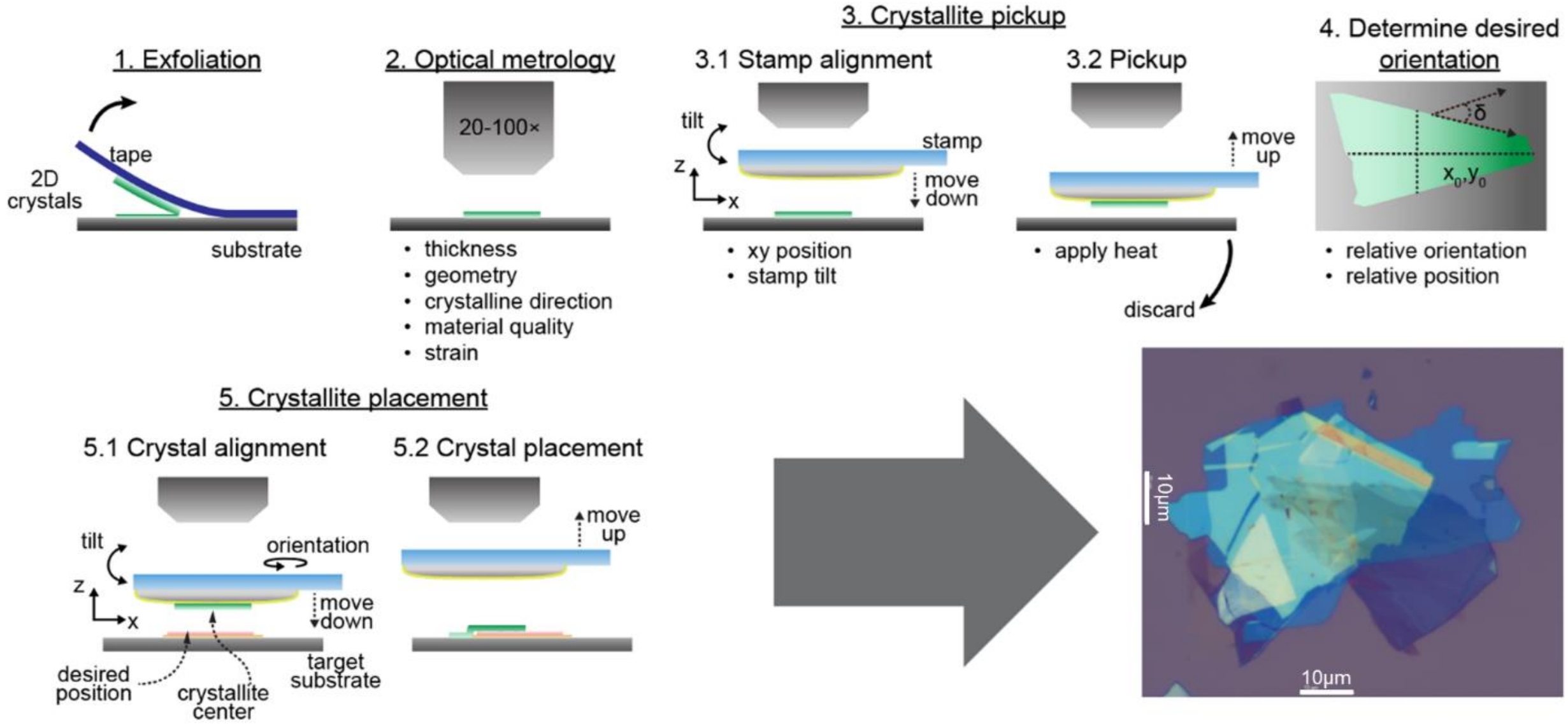

**Figure 1 – general workflow for the layer-by-layer assembly vertical heterostructures of van der Waals materials.** As discussed in the main text, this general recipe results in clean interfaces for a two-layer system, but will suffer from significant polymer contamination for heterostructures with more layers.

This assembly process is wrought with potential pitfalls. First, producing high-quality, minimally contaminated 2D crystallites by mechanical exfoliation is challenging, with production rates that are hindered by the need to sort through hundreds of exfoliated crystallites to confidently identify those with desired thickness and geometry. This production rate is further slowed if an accurate measurement of the crystallographic orientation is needed. Second, polymer materials are commonly involved in both exfoliation and stacking processes. Any time these polymer-based materials contact a 2D crystallite, the contacted surface will be contaminated with remnant polymers that disrupt interfacial coupling and introduce disorder. Third, most transfer systems rely on manual micromanipulation of the stamp for both positioning and stamping. As a result, each stacking step is marred by significant risks of misaligning/misorienting the flake as well as fracturing or tearing either the crystallite being stacked or the partially assembled heterostructure on which it is to be stacked. Despite these challenges, substantial investments of creativity, innovation, and perseverance, mostly by talented graduate and undergraduate researchers, have enabled significant scientific discoveries based on the manual assembly of 2D heterostructures. Thanks to this effort, the community now has a quiver of procedures for the manual assembly of 2D heterostructures that have been covered in detail in other review articles over the past six years[26-33] and will only be discussed briefly here.

*2.1 – 2D crystallite transfer: wet versus dry processes*

For the assembly of 2D heterostructures, the *transfer* process of a 2D crystallite from a host substrate to a target substrate is typically categorized as either a wet process or a dry process. Wet processes are typically employed for chemical vapor deposition- (CVD-) grown crystallites and use a solvent to delaminate the 2D crystallite from its host substrate. Following the delamination, the 2D crystallite is "fished" from the solution using the target (or intermediate) substrate. Several variants of wet-transfer techniques exist that utilize different combinations of support materials, host substrate surfaces, and solvents, and readers are referred to Schranghamer *et al.[28]* for an in-depth review of each one. For 2D heterostructure fabrication, wet-transfer processes expose the 2D material to chemical solutions, which lead to substantial contamination, limiting their utility for high-quality multilayer heterostructures. Further, the fishing procedure is not ideal for the precise alignment/orienting of multiple layers.

The stricter requirements for high-quality 2D heterostructures usually require dry-transfer processes that eliminate exposure of the 2D layers to solvents. For these processes, picking up and placing of the 2D crystallites is accomplished with a solid stamp, which is commonly composed of micromolded polydimethylsiloxane (PDMS) that is coated with a thin layer of a thermoplastic such as polycarbonate (PC) or polypropylene carbonate (PPC)[2,34-37]. By tuning the temperature of the thermoplastic film to either below, at, or above its glass transition temperature, the adhesion and mechanical compliance of the stamp can be tuned. Further, the thermal expansion or contraction of the stamp can be leveraged to bring the stamp into and out of contact with a substrate more gently than manually-controlled mechanical motion.

A significant challenge for dry transfer processes is contamination of the 2D crystallites with residue from the thermoplastic stamp. Many procedures incorporate cleaning steps to partially remove the contamination. However, as with the wet-transfer process, the exposure of the 2D crystallite to solvents adds another possible source of contamination. Further, most cleaning processes fail to completely remove the residual polymers. Therefore, it is generally accepted that the best (and only) way to achieve the most pristine interfaces between adjacent layers is to ensure that the respective surfaces never come into contact with a polymer material. Luckily, the process of mechanical exfoliation produces relatively clean surfaces because both were most likely freshly cleaved from a thicker crystal. Picking up this first mechanically exfoliated crystallite contaminates the top surface with the stamp, but the bottom surface is still relatively clean. The crystallite on the stamp can then be placed onto another freshly exfoliated crystal with a nearly pristine surface to form a high-quality 2D heterostructure of two layers with one good interface[36]. However, following this approach, where the heterostructure is assembled on a target substrate, inclusion of additional layers will be plagued by highly contaminated interfaces, limiting its applicability to more complex systems.

*2.2 – Assembly of many-layer twisted 2D heterostructures with the van der Waals pickup method*

To assemble 2D heterostructures with $N$>2 layers and high-quality interlayer interfaces, the only viable process is the van der Waals pickup method[2,38]. As summarized in Figure 2, the key to this approach is to assemble the heterostructure layer-by-layer on the stamp itself. First, a thermoplastic stamp picks up what will be the top layer of the heterostructure. As discussed above, the bottom surface of this crystallite is not contaminated with polymer material, and it is then lowered onto the next layer of the heterostructure. However, rather than releasing the 2D crystallite from the stamp, the van der Waals adhesion between the two 2D crystallites is used to pick up the second layer, creating a two-layer heterostructure on the stamp. This process is repeated for all subsequent layers of the 2D heterostructure, minimizing interfacial contamination by using surfaces that have not directly contacted polymer material. In these assembly processes, the temperature of the stamp is varied as discussed above, sometimes to temperatures that are well above the glass transition of the polymer to promote the van der Waals adhesion between the layers of the stack[35].

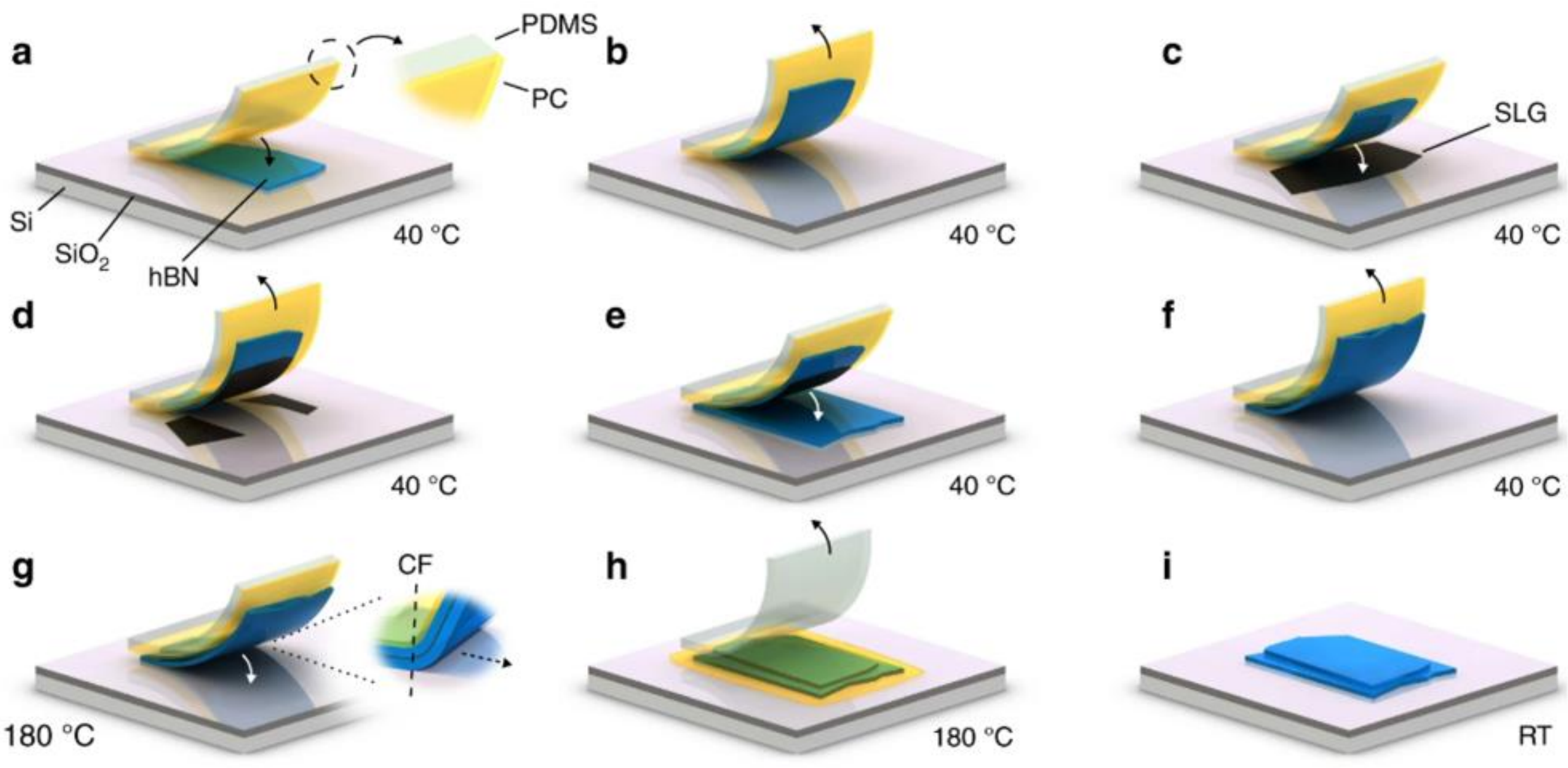

**Figure 2 – Assembly of vertical heterostructures of 2D materials using the van der Waals pick-up method.** (a), (b) a polymer stamp is used to pick up a first 2D layer, which is typically hBN. This first layer will typically be the top layer of the heterostructure. (c)-(f) Subsequent layers are picked up from substrates and assembled on the stamp. (g)-(i) The assembled heterostructure is removed from the stamp by placing it on a target substrate. Residual polymer material from the stamp is then removed from the system by washing the final assembly with organic solvents. Figure reproduced from ref. [38] with permission.

Another important note is that achieving precise interlayer twists when the adjacent layers are the same material (i.e., a homobilayer) can be accomplished with the "tear-and-stack" method, where a single 2D crystallite is cleaved (by tearing or laser cutting) into two or more parts. Each part can then be successively picked up and precisely rotated relative to the other[37], achieving precise relative rotational orientations between the layers. Alternatively, polarized Raman[39-42] and second-harmonic generation (SHG)[43-46] can be used to determine the crystalline orientation of individual crystallites. However, these approaches require precise alignment of the polarization states of the spectroscopy measurements to the geometric features (e.g., edges) of the crystallites, which introduces additional uncertainty in the interlayer twist. Further, the relationship between the crystalline structure and any observed polarization anisotropy will be material (and strain) dependent. As a last resort, if neither the tearing and stacking nor polarization-dependent spectroscopies are available, straight edges of the crystallites, along with the crystal structure, can be used to infer the crystallographic orientation from the optical image of the crystallite itself[47].

Once the 2D heterostructure is assembled on the stamp, it can be deposited onto a final surface through a standard release recipe of lowering the heterostructure-laden stamp onto the target substrate, heating the stamp to a temperature above its glass transition, and then lifting the stamp from the substrate. Because of the heat, the adhesion between the 2D heterostructure and the stamp is weaker than the van der Waals adhesion of the heterostructure to the substrate. After this step, a significant amount of polymer material remains on the target substrate, which can be cleaned using a solvent wash. Because it is already assembled, this final solvent wash is not expected to significantly degrade the interlayer coupling of the 2D heterostructure. A similar rationale applies to the final steps of lithographically fabricating any electrodes that are needed to electrically address the heterostructure. Resist deposition, writing, development, metal deposition, and lift-off are presumed to also have little impact on the performance of the 2D heterostructure. In many cases, final annealing or mechanical squeegeeing steps[26,48] are used to further promote the interlayer coupling and remove top layers of polymer materials.

While the van der Waals pick-up approach has been successfully utilized for several high-quality 2D heterostructures and groundbreaking scientific advances[2,49-55], efforts are being made to improve the quality of the heterostructures produced, especially by eliminating contamination. Jin *et al.* introduced a novel dry assembly technique that eliminated both polymers and solvents for creating van der Waals heterostructures with pristine interfaces[56]. The effectiveness of their approach was validated through ambient atomic force microscopy (AFM) and atomically resolved scanning tunneling microscopy (STM), confirming the production of air-sensitive heterostructures with ultra-clean interfaces and surfaces. Their technique could be compatible with UHV conditions, offering the potential for the highest quality heterostructures. In 2023, Wang *et al.* introduced a novel polymer-free method to assemble van der Waals heterostructures that utilizes flexible silicon nitride membranes as transfer supports to similarly eliminate the need for polymeric materials[57]. The key advantages of this method are its high-temperature compatibility and its versatile compatibility with different environments. This method is explained in more detail in Section 8.

*2.3 – Improving 2D heterostructure assembly with automation*

Despite these advances and success stories, the 2D heterostructure fabrication process remains very cumbersome. There is still substantial room for improvements and innovations that increase its versatility, improve compatibility with air- and solvent-sensitive materials, increase the yield and rate of the assembly process, further decrease contamination, and enable the pursuit of more complex multi-layered 2D heterostructures that are compatible with increasingly sophisticated measurement techniques such as angle-resolved photoemission spectroscopy (ARPES), terahertz STM, and nano-optical spectroscopies.

To date, the vast majority of 2D heterostructure assembly—from exfoliation to crystallite identification, to 2D assembly—has been executed manually, where a human operator is responsible for performing and monitoring all of the steps of each process either directly or through motorized motion controllers. However, analogs of all of these operations have been automated either in fabrication facilities or robotic laboratories, where advanced image recognition algorithms, machine learning, and AI are synergized with robotic automation for substantially greater control/precision, longer duty cycles, and better reproducibility than what can be achieved with a human operator (who gets tired, distracted, frustrated, bored, etc.). Adapting these automation tools for 2D heterostructure fabrication can ease the burden of sample preparation as well as improve the quality and reliability of the samples that drive scientific innovation. This potential of robotic automation is increasingly recognized as a fruitful avenue to both address reproducibility issues[32,58-60] as well as pursue more complex multi-layered heterostructures where the odds of successful fabrication are prohibitively low if using manual processes[61,62].

## 3. Automated exfoliation of 2D materials

Mechanical exfoliation of bulk van der Waals crystals is the most versatile approach for producing individual single-layer (and few-layer) crystallites of 2D materials. In principle, it works for all van der Waals materials, it can be performed at a relatively low cost, and it can be implemented in a variety of settings from a simple desk under ambient atmosphere to a robotic system in an UHV chamber. Extensive reviews of manual mechanical exfoliation of 2D materials can be found elsewhere[63] and cover major advances, including gold[64] and gel[65] exfoliation processes. Currently, there are two major drivers for automating mechanical exfoliation. The first is the strong, presumably ubiquitous desire of 2D materials experimentalists to improve the yield, quality, production rate, and/or size of the single-layer crystallites produced by the mechanical exfoliation process. Manual exfoliation methods suffer from inconsistent flake quality, low yield, and limited scalability. The second is to be able to exfoliate 2D materials in UHV environments to minimize contamination and exposure to ambient atmosphere. In this section of this review, we focus on the development and adaptation of robotics to automate and improve the mechanical exfoliation of 2D materials and perform the process in UHV.

### *3.1 – Advantages of Automated Exfoliation for 2D Materials*

Robotic automation offers many potential advantages for improving the output of mechanical exfoliation. First, modern motion control systems, as typically found on 3D printers, computer numerical control (CNC) machines, and automated industrial fabrication lines, routinely synchronize complex movements with precisions that reach the scale of tens of nanometers. This movement control dwarfs what is achievable with human-driven manual processes and can be leveraged to control parameters such as applied forces, rates, and temperatures with exceptional precision. Because these settings can be recorded and repeated, results can be reproduced across different labs that have similarly capable instruments. Additionally, robotic systems can repeat any recipe with unparalleled reproducibility and have nearly continuous duty cycles, which enable brute-force searches for optimum process parameters that are otherwise far too tedious to conduct manually. Plus, even if such optimization and precision motion control ultimately only marginally improve 2D crystallite production, automated exfoliation systems are able to run continuously, producing high volumes of crystallites where even small production yields will result in relatively large numbers of usable samples for research. By lowering dependence on operator skill and allowing systematic tuning of parameters, robotic exfoliation is a crucial step towards scalable and standardized fabrication of layered heterostructures.

*3.2 – Advances in Automated Exfoliation Techniques*

Despite the potential advantages, only a few groups have reported formal efforts to automate the exfoliation process. These apparatuses are summarized in Figure 3. DiCamillo *et al.* retrofitted a rheometer—an instrument that normally measures the flow of liquids or soft solids under force—to successively mechanically exfoliate $MoS_2$ and $MoTe_2$ crystals between two adhesive surfaces (Nitto blue tape), automating the "copy-and-paste" process of exfoliation (Figure 3c)[66]. Using this apparatus, the authors tested the effects of the number of copy-and-paste operations as well as the force applied during each operation. Following the automated copy-and-paste operation, the authors manually transferred crystallites from the tape onto $SiO_2$/Si chips. Coarse analysis of the resulting crystallites confirmed that average thickness decreases with increasing numbers of copy-and-paste operations and revealed that greater force used in the copy-and-paste operations may increase the likelihood of producing single- and few-layer crystallites. More recently, Kobayashi *et al.* presented a simple apparatus from off-the-shelf parts (Figures 3e and 3f) that automates a peeling operation to transfer (and exfoliate) 2D materials affixed to adhesive tape onto a target substrate[67]. They demonstrated control of both the rate and angle at which the tape was peeled from the target substrate. With their apparatus, the peeling angle can be set anywhere between 0 – 90 degrees (w.r.t. the normal of the substrate) and the accessible peeling rate spans two orders of magnitude. They demonstrated that both the slowest exfoliation speed of 0.014 cm/min and a peeling angle of 60 degrees resulted in the greatest average areas of single-layer crystallites, demonstrating the potential for optimization.

Courtney *et al.* developed the "eXfoliator" instrument that precisely controls the substrate temperature, the peeling rate (or velocity), the peeling angle, and the pressure at which the tape is applied to the substrate[68]. Over a detailed study of 33 runs, the authors concluded that several other process parameters such as the loading of the tape with 2D materials, the ambient humidity, and contamination on the destination substrate are as important as the precisely controlled parameters of their instrument. Nevertheless, they demonstrated that the eXfoliator reliably generates more than three crystallites of large-area graphene every exfoliation run. Gasbarro *et al.* recently reported a simple, very cost-effective apparatus designed for carefully testing the effects of peeling angle and peeling speed[69]. They employed a changeable mount that fixes the peeling angle and allows for a very broad range of peeling angles to be explored. Their initial report validated the performance of the instrument and provided a detailed bill of materials, making it very easy to be adapted by other groups across the world. Sozen *et al.* invented a reel-to-reel deposition tool for high-throughput fabrication of continuous 2D crystallite films[70]. The apparatus prioritizes large-area coatings and does not control the thickness of the 2D materials deposited, limiting its applicability to producing materials in the single- and few-layer regimes. However, the high-throughput nature of the reel-to-reel design offers inspiration, especially for maximizing throughput. And finally, researchers at Brookhaven National Laboratory have pioneered a roller-based exfoliation approach in their QPress system[71-78].

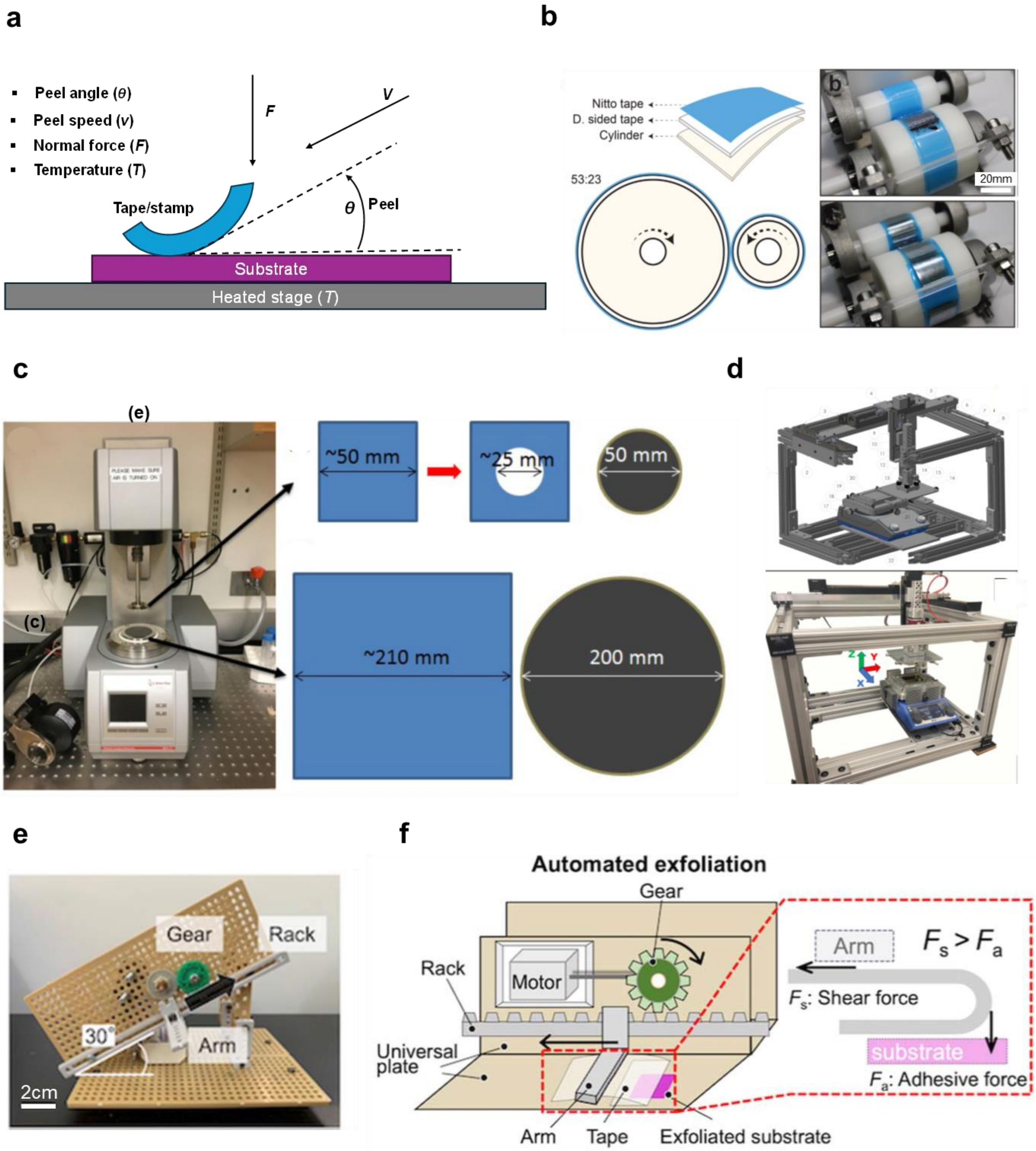

**Figure 3 – Apparatuses for mechanized/robotic exfoliation.** (a) Parameters controlled during robotic exfoliation: peel angle (θ), peel speed (v), normal force (F), substrate temperature (T) Robotic exfoliation apparatuses. (b) Schematic and photos of the massive parallel exfoliation setup using two polyoxymethylene cylinders (c) Rheometer-based exfoliation setup using a 50 mm diameter tool and blue tape affixed with a circular cutout, alongside a 200 mm stage and correspondingly sized blue tape. (d) Design and implementation of the eXfoliator device. (e) Simplified rack-and-pinion setup illustrating the basic mechanical exfoliation principle. (f) An automated exfoliation schematic. (b) reproduced from ref. [70] with permission. (c) reproduced from ref. [66] with permission. (d) reproduced from ref. [68] with permission. (e), (f) reproduced from ref. [67] with permission.

All of the reported exfoliation apparatuses could operate in a glovebox under inert atmosphere. However, the ultimate environment to minimize degradation and contamination is UHV. Several solutions for vacuum exfoliation have recently emerged[79-82], achieving highly pristine samples that meet the needs of the most demanding experiments. Metal films combined with the pristine UHV environment result in the production of single layers that are limited in size only by the size of the bulk crystal. Scaling these techniques for high-throughput production of 2D materials will be challenging and will likely require solutions that adapt several techniques of the apparatuses built for exfoliation in ambient/glovebox atmospheres to work in UHV environments.

Finally, liquid-phase exfoliation is an alternative to tape-based methods for producing 2D materials. This approach involves mixing bulk materials in solvents and then using ultrasonication to break them down into thinner flakes, typically producing crystallites that are sub-micron in size. Nevertheless, the process can be automated to precisely control factors like the temperature, sonication time, and how much solvent is used, ensuring the exfoliation results are consistent. In automated systems, steps like filtration and centrifugation are used to separate the flakes by their size and thickness. In 2014, Coleman and his team investigated a shear-based method of liquid-phase exfoliation, showcasing how varying sonication time and type of solvent can influence the quality and quantity of graphene produced. The study notably shows the potential for scaling up the process for industrial applications, emphasizing its viability for mass production of high-quality graphene[83].

Liquid-phase exfoliation generates large amounts of nanosheets in solutions, but often at the cost of flake quality. Processes such as ultrasonication and shear mixing tend to create edge defects and disrupt the crystal structure, which can reduce electrical and optical performance. Surfactants or solvents used to keep the flakes suspended may also remain on their surfaces, leading to contamination that is difficult to remove. Additionally, the resulting flakes usually have a wide range of thicknesses and sizes, making it hard to isolate clean, uniform monolayers. Because of these challenges, liquid-phase exfoliation is useful for producing inks or composite materials, but it is generally not suitable for high-performance or device-grade heterostructure fabrication.

*3.3 - Outlook*

Most automated exfoliation systems strive to control the same key parameters, which include the angle of peeling, the applied force, and the temperature. Although they vary widely in how these parameters are controlled, there are promising indications that the control can improve the quality and yield of lab-scale production of 2D crystallites for scientific research, which is a significant step that needs to be taken on the path towards industrial-scale manufacturing. Systems like the eXfoliator[68] and the configurable gear and rack system[67] by Kobayashi *et al.* show that actively controlling the peeling speed can improve both the yield and consistency of produced 2D crystallites. In contrast, high-throughput reel-to-reel systems focus on processing speed and scalability, at the expense of quality and consistency of the 2D crystallites, leaving a major gap between large-scale production and production of high-quality crystallites. In addition, none of these systems yet implement adaptive feedback that can adjust the exfoliation conditions in real time based on how each flake is responding mechanically. Future improvements may be found with the integration of mechanical control with real-time feedback from optical or force sensors, allowing the peeling conditions to automatically adjust based on adhesion strength or flake thickness. When combined with AI-driven optimization, this approach can make exfoliation more consistent and less reliant on the skill of the operator, making it key to scaling the process beyond research laboratories.

## 4. Automated flake detection and metrology using artificial intelligence

Mechanical exfoliation produces an ensemble of crystallites of different thicknesses from which crystallites of the desired thickness must be identified and extracted. This cataloging process is accomplished with optical microscopy, which is prime for robotic automation and image processing. Automatic optical detection and subsequent classification (i.e., prediction of thickness) of 2D crystallites usually rely on brightfield reflectance microscopy, where contrast and RGB values typically serve as the identifiers of 2D materials (Figure 4a). For example, Figure 4b illustrates the distinct contrast clusters formed by graphene in color contrast space. More advanced imaging modalities that use the reflectance and transmission spectra of 2D materials, such as elastic hyperspectral imaging[84-87] or spectroscopic imaging ellipsometry[88], are less common but gaining increasing traction. Automating the collection of imagery data of substrates with 2D crystallites is relatively routine thanks to the commercial development of automated microscopes for biological research and semiconductor device inspection.

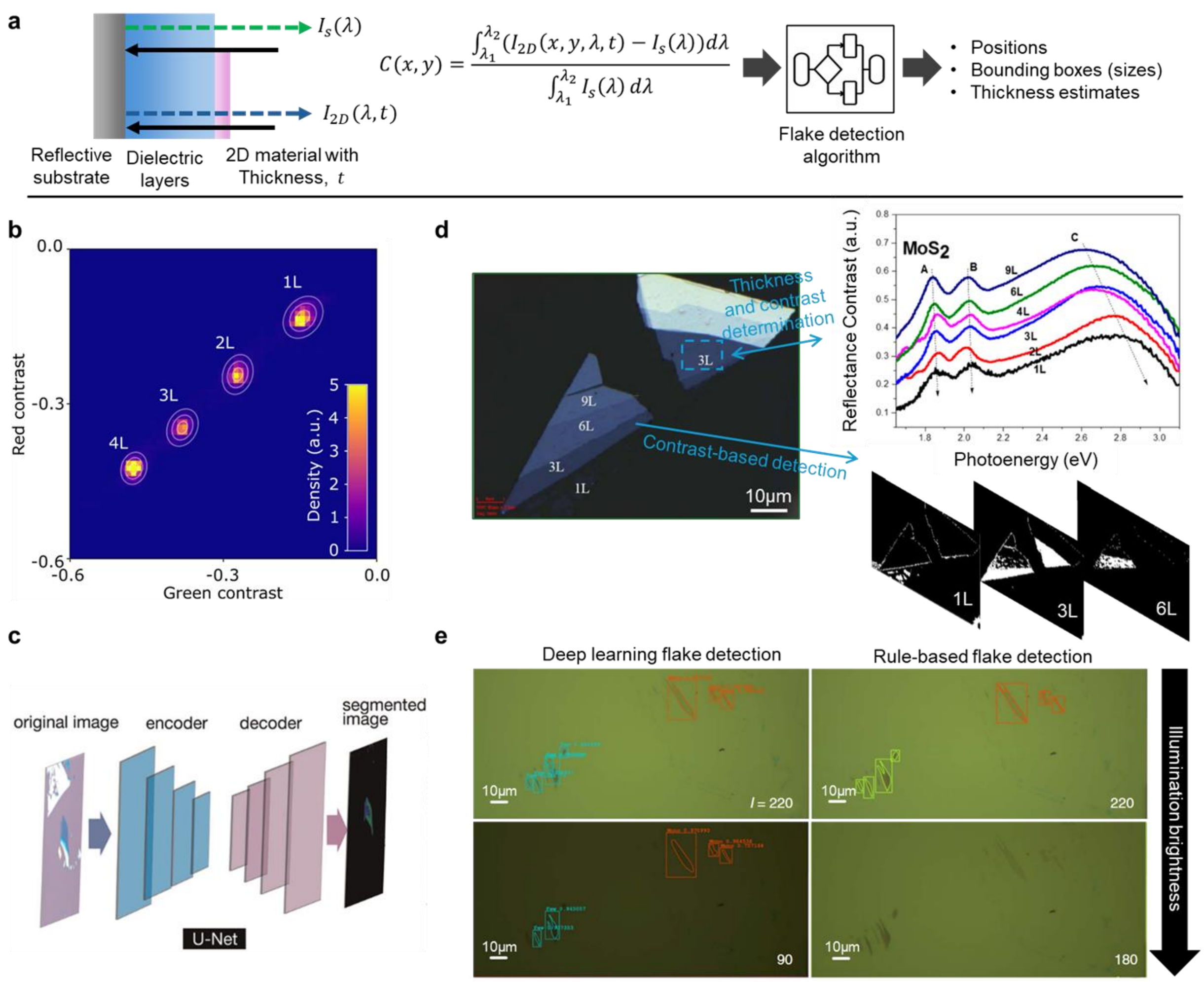

**Figure 4 – Computerized optical identification and metrology of 2D crystallites.** (a) Schematic showing the origin of the optical contrast of 2D crystallites. The reflectance of the substrate is modified by the presence of a 2D crystallite through thin film interference. The contrast is a function of wavelength and 2D crystallite thickness, enabling the distinction of crystallites of different thicknesses. (b) RG color space contrast values for graphene on 90nm SiO2/Si. The isolated clusters allow for layer identification. Panel from ref. [89]. (c) Schematic of a U-NET based neural network identifying flakes with different layer numbers. Panel from ref. [90]. (d) Schematic of hyperspectral-assisted contrast detection. Differential reflectance spectroscopy is used to determine the contrast corresponding to material of a specific thickness. Then this contrast is used to detect all other material of the same thickness. Panel adapted from ref. [86]. (e) Comparison of neural network and threshold (rule)-based detection of 2D materials for images of different brightness. The neural network continues to produce accurate predictions even when the brightness is reduced by more than a factor of 2×, while the threshold-based detection already fails when the brightness is reduced by ~20%. Panel from ref. [91]

### *4.1 – Automatic detection and analysis of optical imagery of 2D crystallites*

A greater challenge is the automatic processing of the large volumes of collected data sufficiently fast to avoid slowing down the data acquisition. The software should be robust to perturbations to the imaging conditions and rapidly and accurately identify high-value 2D crystallites, which can range from monolayers to crystallites with a specific layer number in each image. For the image processing, the two main approaches are either an AI-based machine vision algorithm[77,85,87,89-110] such as U-Net (Figure 4c), or a more traditional image processing algorithm that uses threshold values to detect 2D crystallites in the images[68,84,86,88,111-117]. The advantages of the second approach are the relative ease of implementation, the small number of images needed for development[112-115], and detection accuracies approaching or exceeding 90%[112,113,115]. However, these methods are sensitive to the imaging conditions[112-114,116], require transparent substrates[117], or are material-specific[115]. Therefore, they need to be calibrated for specific imaging systems and different materials. The sensitivity to the imaging conditions can be eliminated if the detection threshold values are based on hyperspectral information[84,86] as illustrated in Figure 4d. There has also been an effort to develop a detection metric based on RGB and RAW pixel values independent of the imaging conditions[111].

The challenges confronting traditional image processing algorithms are circumvented by deep neural network-based detection schemes since the inherent generalization ability renders the detection scheme insensitive to variations in the imaging conditions[91,92,96,97,100,106,118]. This robustness is shown in Figure 4e, with the neural network maintaining identification accuracy over a wide range of brightness conditions compared to a traditional image processing approach. Meanwhile, transfer learning can extend the detection schemes to materials that the original neural network was not trained on directly[91,92,96,97,105,110,118]. These advantages are combined with detection accuracies that can be close to 100%[91,95-97,103,106]. However, the stronger performance comes at the cost of needing a large volume of training data[101,107], high computational cost[101], and an unintelligible crystallite identification process that has been shown, for at least one case, to use physically irrelevant image features as identifiers[107]. These disadvantages can be addressed by using tree-based machine-learning algorithms, at the cost of detection accuracy[101,107].

*4.2 – Integration of Automatic Flake Detection with Motorized Microscopes*

To fully automate the detection workflow of 2D crystallites, the automatic processing of image data must be integrated with a fully automated scanning microscope into a cohesive workflow. Such systems have been reported with neural networks that can identify various 2D transition metal dichalcogenides (TMDs)[89,91], graphene[77,89,91,108], and hBN[77,91,108]. Also, the integration of an empirical RGB value-based detection scheme for graphene[68] and an empirical contrast-based detection scheme for TMDs[112] with scanning microscopes have been achieved. Additionally, the integration of a detection scheme using hyperspectral imaging to identify the thickness of TMDs, which is then used to set contrast bounds for the detection of flakes with the same thickness in brightfield images with a scanning microscope, has also been reported[86]. These systems can reliably identify 2D crystallites of interest, achieving accurate identification rates of 60-97%[77,89,91,108,112], depending on the material, flake size, and performance specifics of the detection scheme used. Meanwhile, the mechanical performance and throughput of these systems vary greatly. The volume of substrate chips that can be scanned during a single acquisition run varies from one[86,108] to nine[77] to however many can be put on a sample tray[89] to a whole wafer[68]. Likewise, the overall time of the data acquisition workflow also varies widely. Uslu *et al*. estimate that the high-resolution scan speed of their instrument is 25 $cm^2$/hr, while Courtney *et al*. estimate their data acquisition time for a whole wafer is 60 minutes. Greplova *et al*. report that their instrument takes 190 seconds to scan one chip. Furthermore, except for the system developed by Chang *et al.*, only brightfield image data is used, neglecting information on the number of layers contained in spectral data.

Future development efforts should focus on integrating the most successful detection schemes with high-throughput automated microscopes capable of collecting several types of spectral data along with brightfield optical images to allow for the unambiguous identification of 2D crystallites. For instance, the QPress system at Brookhaven National Lab system[119] integrates Raman and photoluminescence, both of which can greatly increase the confidence and accuracy of crystallite classification.

### *4.3 – Improving the Visibility of Low-Contrast Materials*

There have also been efforts to make low-contrast materials, such as hBN, visible in optical images, which would allow for their automated detection. Engineering the optical properties of the hBN/substrate stack has proven to be an effective method to achieve this requirement. Applying a thin-film coating of SiN to Si allowed for the reliable identification of monolayer hBN in optical images using white-light illumination[120]. The same effect was achieved with a specially designed Si/SiN/$SiO_2$ substrate, with further contrast enhancement if a band-pass filter isolates the blue part of the illumination spectrum[121]. Similar contrast enhancement was observed using a thin gold layer deposited onto Si/$SiO_2$[122]. This improvement can be seen in Figures 5a and 5b, where the thin hBN is clearly visible on both types of substrate stacks. Meanwhile, thin AlN films have been shown to enhance the contrast of both graphene and $MoS_2$[123]. Also, using the wavelength shift of reflectance minima in addition to an engineered substrate stack has been used to develop a method of layer identification for graphene and TMDs suitable for wafer-scale processes important for future industrial applications[124]. This process was extended to determine the number of layers and composition of van der Waals heterostructures[125], potentially allowing the automatic characterization of heterostructures. As an alternative to substrate engineering, using the blue color channel of white-light images has been shown to enhance the contrast of hBN on transparent substrates[126], and illuminating the transparent substrate at its Brewster angle has been used to improve the contrast of graphene[127]. Figure 5c shows the reflectivity plot obtained using the latter technique, with the inset showing that the contrast of graphene is enhanced by several orders of magnitude when the sample is illuminated at the Brewster angle of the substrate.

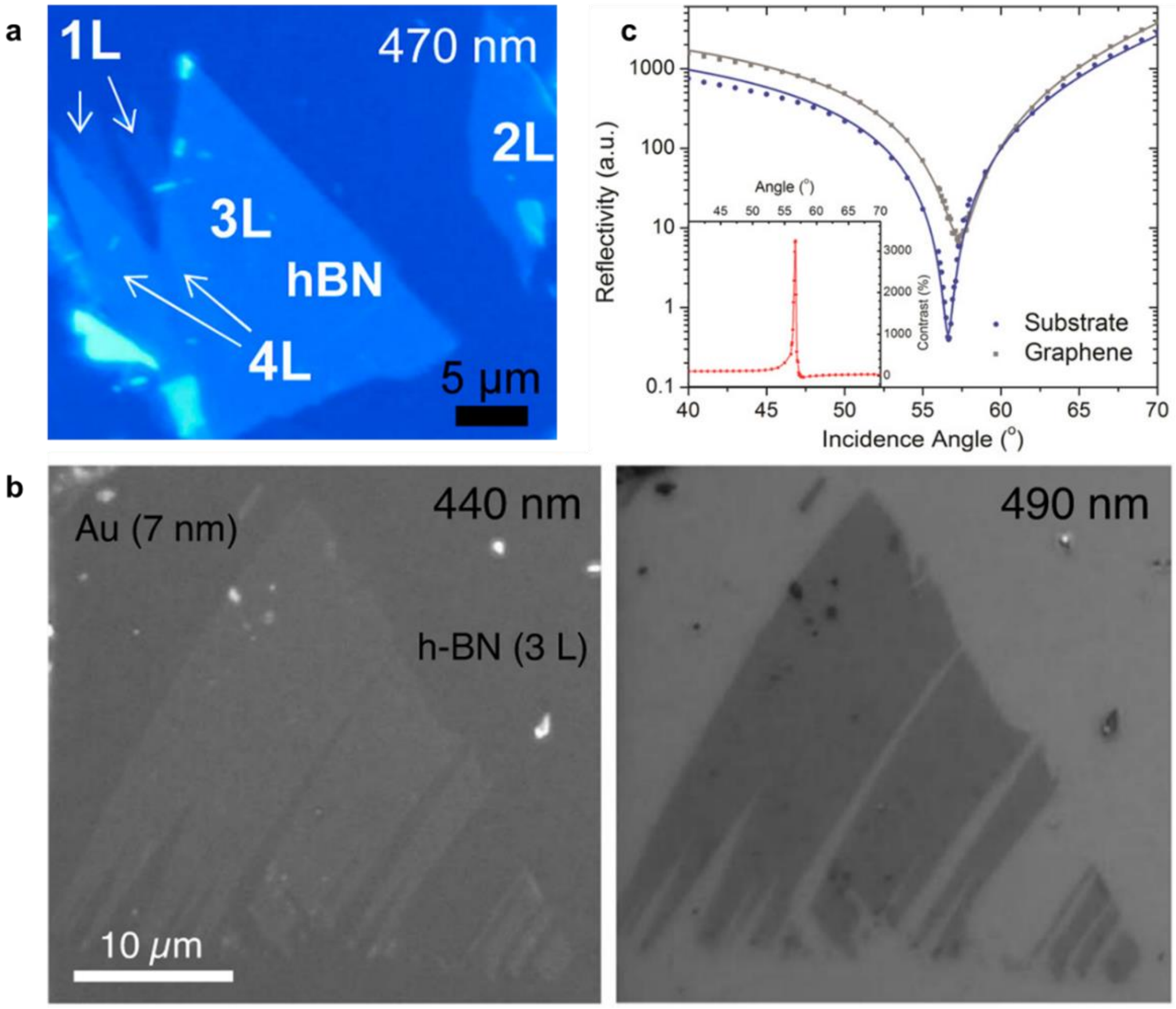

**Figure 5 - improving optical contrast of 2D crystallites.** (a) Image of an hBN flake on a designer Si/SiN/SiO2 stack under blue-light illumination showing that the monolayer is clearly visible. Panel from ref. [121]. (b) Monochrome image of a 3L hBN flake on a designer Si/SiO2/Au stack. Narrow bandpass filters centered at 440nm (left panel) and 490nm (right panel) were used when acquiring the images. Note the significant contrast enhancement of the hBN flake when using the 490nm bandpass filter over the 440nm bandpass filter. Panel from ref. [122]. (c) Reflectivity plot as a function of incidence angle for both the substrate (borosilicate glass) and the substrate/graphene stack. The inset shows the contrast of the graphene relative to the substrate. The large dip (spike) in the reflectivity (contrast) occurs at the substrate's Brewster angle. Panel from ref. [127].

*4.4 – Neural Networks for 2D Material and Heterostructure Metrology*

Finally, neural networks are starting to be applied to more than 2D crystallite identification. They have recently been employed to determine the optical constants of different 2D materials efficiently from reflectance spectra, greatly reducing both the experimental[128] and computational[129] effort usually required. Figure 6a shows the excellent agreement between the neural network-predicted and experimentally extracted optical constants. Being able to efficiently determine the optical constants, especially for novel 2D materials, could aid in the development of physically informed neural networks (PINNs) for crystal identification. Automated morphology characterization by neural networks of 2D crystallites has also been used to rapidly estimate their mechanical and physical properties[110,130]. The successful identification of the material layers in a TMD heterobilayer by a neural network based on brightfield reflectance images has also been reported[105], illustrating the potential to automate some basic characterization of TMD heterostructures. Additionally, neural networks have been used to characterize crystallites grown through CVD. Here, they have been successfully used to determine the material[131], the thickness of $MoS_2$ flakes, and the twist angle of as-grown bilayers[132], as well as the quality of the CVD-grown flakes[133], from brightfield reflectance images. The close agreement between predicted and actual twist angles, as verified by second harmonic generation measurements, obtained in ref. [132] is shown in Figure 6b. Neural networks have also been applied to the identification of layer thickness, impurities, and stacking order in van der Waals heterostructures[104], with Figure 6c showing the successful layer identification in a Graphene/$MoS_2$ heterostructure. Lastly, neural networks have also been used to automate the analysis of twisted bilayer graphene Raman spectra, allowing for the rapid determination of the twist angle[134], potentially paving the way for a new, efficient method for determining the twist angle of van der Waals heterostructures

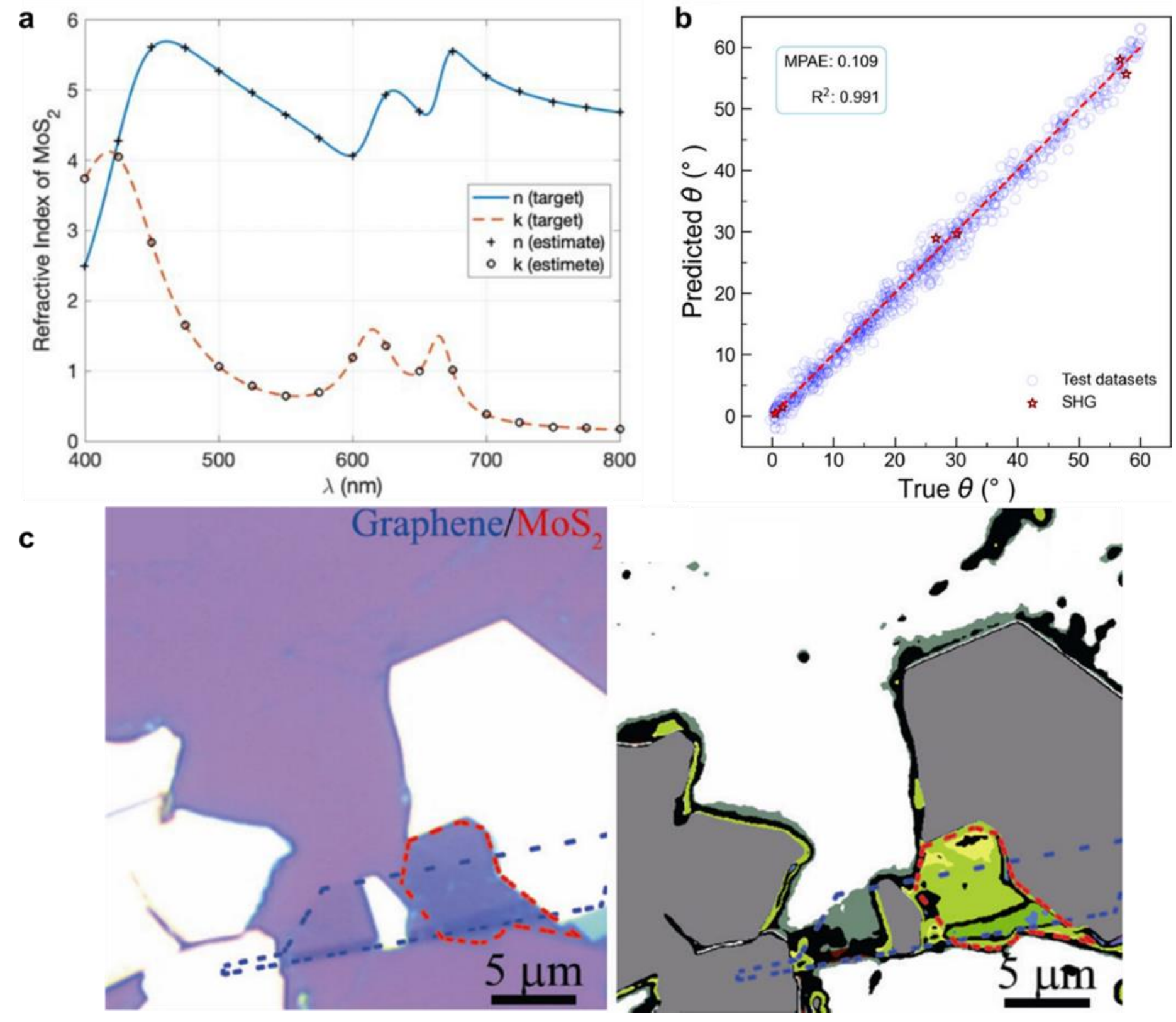

**Figure 6 - AI-assisted metrology of TMDs.** (a) Comparison between the neural network predicted (dots) and experimentally extracted (lines) values of n and k for MoS2. The inputs into the model are differential reflectance spectra, from which n and k are directly predicted. Panel from ref. [129]. (b) Neural network-predicted twist angles vs actual twist angles for a test data set (blue dots) and SHG derived values (red stars). The dashed red line is the desired relationship. Panel from ref. [132]. (c) Optical image (left) and support vector machine (SVM) model-based layer predictions (right). Graphene ($MoS_2$) is marked with a blue (red) outline in both panels. In the right panel, dark green, light green, and teal colors represent the SVM predicted areas of heterostructure, MoS2, and graphene, respectively. Panel from ref. [104].

### *4.5 – Summary and Outlook*

The future is bright for utilizing AI to expedite both the fabrication and characterization of 2D materials and heterostructures. Already, multiple models have been developed to detect 2D crystallites, with some of them having been integrated into motorized scanning microscopes. Such automated optical imaging and machine-learning tools have already reduced the amount of time needed to manually search for suitable flakes. Utilizing substrate engineering to improve the visibility of low-contrast materials could improve the efficiency of the detection models for such materials. Additionally, neural networks have been demonstrated to be accurate in determining the optical constants of 2D materials, as well as performing basic morphological and structural analyses of 2D materials and their heterostructures. Despite this progress, there is still a lack of general access to the detection models since not all publications provide open-source access to their algorithms. Furthermore, work is still needed to make the detection models agnostic to the change in appearance of 2D materials due to variations in the $SiO_2$ thickness. And finally, process optimization, especially for exfoliation, would benefit from the rapid integration of these metrology advancements into robotic systems such as the 2DMMS[60] that unite exfoliation and optical cataloging. In this direction, the 2D materials preparation and fabrication community greatly benefits from openly available software packages such as 2DMatGMM[135] and QuantumFlake[136] that standardize the implementation of crystallite detection.

## 5. Robotic assembly of 2D heterostructures

Although conceptually simple, the assembly/fabrication of high-quality 2D vertical heterostructures is technically challenging. Critically, conventional approaches to heterostructure assembly[31-33,137,138] rely on manual, human-controlled micromanipulation of individual atomically thin layers attached to homemade (polymer and inorganic) stamps, guided by typically unoptimized optical microscopy. As a result, the complexity of the fabrication, the risk of catastrophic failure, the chance of significant interfacial contamination, etc., all dramatically increase with the number of layers that need to be assembled. Analyzing this complexity with simple statistical analysis, Masubuchi *et al*. estimated that over 120 hours of continuous operation would be needed to manually assemble a system with $N = 29$ layers[60]. These fabrication liabilities ultimately hinder large-scale scientific exploration of 2D heterostructures that consist of more than a handful of layers, where a wealth of scientific discoveries may lie[139-143]. Further, the manual fabrication processes are typically deemed too complex, high-risk, and/or time-consuming to reproduce groundbreaking results in the field – an issue that has been explicitly called out in the last few years for 2D moiré systems[144,145]. Sooner or later, the drive of scientific inquiry will surpass what can be accomplished with the manual assembly of vertical heterostructures.

Robotic automation of heterostructure assembly and stacking of 2D materials is a promising pathway to overcoming this inevitable fabrication-induced roadblock. By borrowing innovations that are now routinely used in modern high-tech fabrication facilities, it is possible to build heterostructure assembly robots with motion control on the nanometer scale, rotational control well below a single degree, and unparalleled control of applied forces and temperatures. Further, these control systems can be integrated with modern AI and machine learning for automated operation. Such automation promises higher yields by reducing operator-driven failures and the potential to significantly reduce the amount of time needed to fabricate high-quality, many-layer heterostructures.

*5.1 - Robotic assembly of many-layer 2D vertical heterostructures in ambient and inert atmospheres*

A few major steps forward in the development of robotic instruments for heterostructure assembly have been made in recent years[60,146,147]. In the first seminal demonstration, Masubuchi *et al.* united AI-driven crystallite detection and classification (see below) with a stacking robot that was capable of semi-autonomous assembly of vertical 2D heterostructures within a glovebox of inert atmosphere. They called their system the 2D Materials Manufacturing System (2DMMS; Figure 7a) and demonstrated its capability by assembling a heterostructure composed of 29 total layers that alternated between graphene and hBN (Figure 7d). During the assembly process, the 2DMMS requires human confirmation of the placement of each layer, making it only semi-autonomous. Nevertheless, the 2DMMS is a significant advance in heterostructure assembly, and the authors comprehensively documented the details of the system, including the mechanical components of the robots, the control and user interface software, the image recognition algorithms, a heterostructure CAD package, and the implementations of backend databases and software workflows that process and organize the data. The 2DMMS is an exemplary open-source foundation for further innovations in robotic heterostructure assembly systems. However, there has only been limited advancement beyond the 2DMMS system, leaving significant room for further process innovation and automation improvement. The Quantum Press (QPress) at Brookhaven National Laboratory (Figure 7b) uses a hexapod to control the orientation and tilt of the stamp, enabling the robot to reproduce a carefully tuned pressing motion of the stamp to "squeeze" out interfacial contaminants. In addition, Ahn *et al.* reported a new stamping procedure that could further increase the reproducibility of the instruments[147], and in a follow-up study, Masubuchi *et al.* reported a new approach to dry transfer that produces heterostructures that are suitable for ARPES measurements[148].

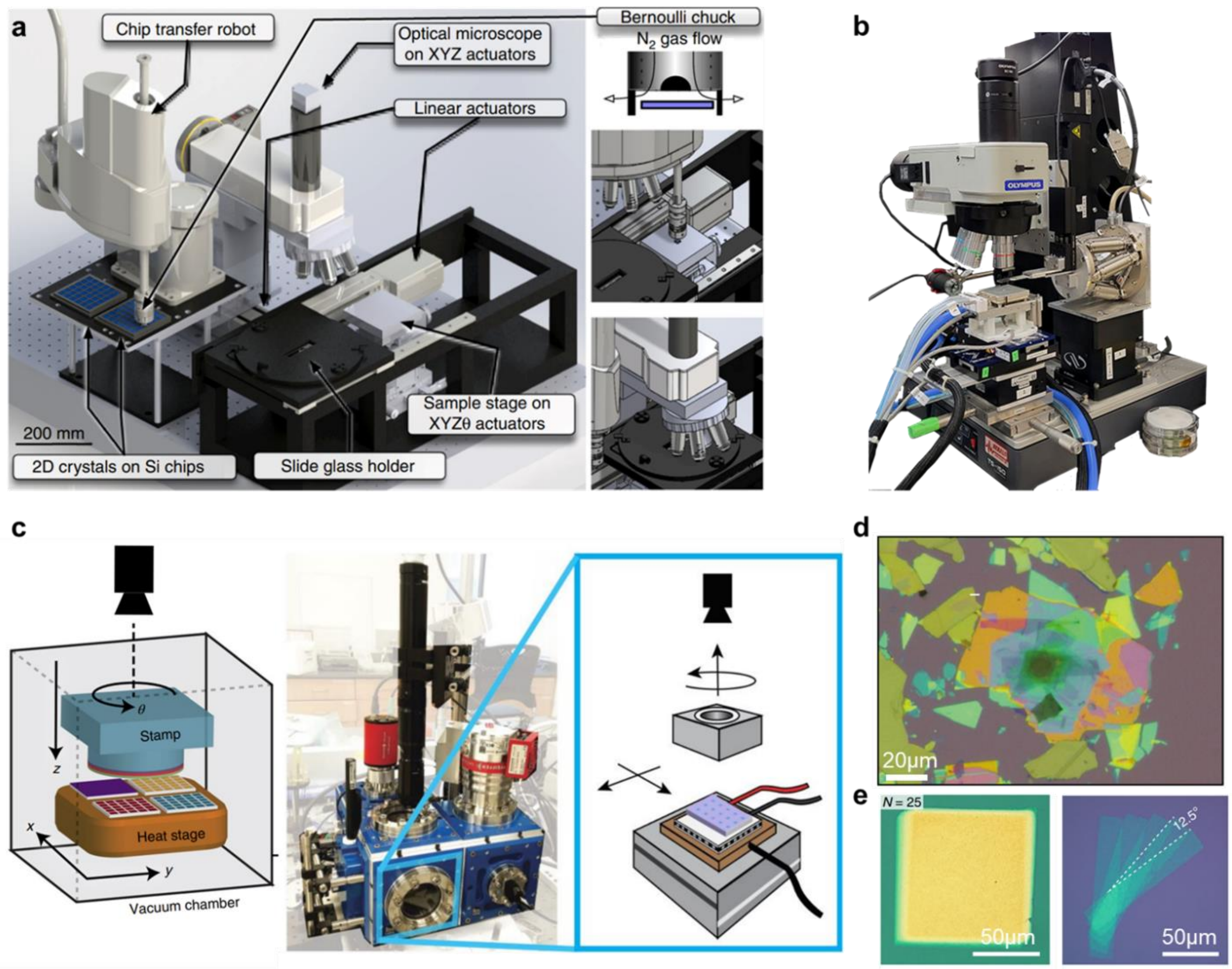

**Figure 7 – Robotic apparatuses for the assembly of 2D vertical heterostructures.** (a) 3D CAD models of the stacker robot of the 2DMMS, including a Bernoulli chuck for manipulating individual chips without making physical contact (right of the panel). (b) Photo (background removed) of the robotic stacker used in the QPress facilities at Brookhaven National Lab. (c) The robotic stacking apparatus built by Mannix et al. for the assembly of heterostructures of CVD-grown 2D materials in a vacuum chamber. (d) A heterostructure composed of $N$=29 alternating layers of hBN and graphite produced with the 2DMMS. (e) Example 2D heterostructure pixels created by the robotic stacking system. (a),(d) Reproduced with permission from ref. [60] with permission. (b) Adapted from the web site for the QPress system [119]. (c),(e) Reproduced from ref. [146] with permission.

*5.2 – Robotics for remote assembly of heterostructures in vacuum*

A vacuum environment is ideal for minimizing contamination during heterostructure assembly (see sections below), but requires some form of mechanized/robotic instrumentation. Mannix *et al.* reported a robotic heterostructure fabrication system that seamlessly operates in a vacuum environment (Figures 7c and 7e). They termed their system the Vacuum Assembly Robot (VAR)[146]. The VAR relies on the prefabrication of "pixels" of 2D materials that are etched from waferscale CVD-grown 2D materials. Importantly, the authors reported a process that created these pixels without contamination, allowing them to construct heterostructures with strong interfacial coupling, as demonstrated by observing lattice reconstruction effects in a twisted TMD heterostructure. Almost certainly, strong interfacial coupling is promoted by operating the system in a vacuum chamber. In addition, their vacuum stacker successfully created multilayer assemblies with up to 80 layers, using an operator-free process at a rate of up to 30 layers per hour. The VAR demonstrated the scientific potential of the large $N$-layer systems by revealing an unconventional layer-dependent progression of the excitonic absorption of $MoS_2$ from 1 to 17 layers that are incrementally twisted as they are stacked. The VAR approach exemplifies how scalable manufacturing can be achieved with 2D materials, and recent advances in wafer-scale growth of single crystals of 2D materials increasingly alleviate concerns about the quality of CVD-grown materials for an increasingly large library of 2D systems[149].

*5.3 – Robotic assembly of 2D materials beyond single crystallites for scalable advanced functional materials*

There have also been advancements in the automated assembly of multilayer 2D material systems for applications beyond the studies of emergent fundamental optoelectronic and magnetic phenomena in 2D heterostructures. Similar to the VAR system, Chen *et al.* reported a combination of pixel fabrication and robotic assembly for the assembly of structural superlubricity applications from HOPG[150]. They demonstrated the ability to assemble 100 "pads" of HOPG in arbitrary locations and patterns in 100 minutes. Their instrument also automatically tested each pad structure to confirm whether or not it possessed structural superlubricity properties. Furthermore, they also leveraged convolutional neural networks at all stages to automate the testing of each HOPG pad, confirm the pickup of a pad, and then place the pad at a given position and orientation. Han *et al.* built a small robotic apparatus that enabled the automated stacking of vertical assemblies of multi-layer vertically-aligned TMD films composed of nanoscale crystallites packed into a single layer, but on edge, such that the interfacial coupling is maximum within the TMD film[151]. While this material system is limited for the study of emergent phenomena in 2D heterostructures because the vertically aligned TMD films do not promote strong interlayer coupling, it represents a facile, scalable fabrication process for optoelectronic applications where the interfacial coupling between the layers is less critical.

*5.4 – Summary and outlook*

Robotic automation of pick-and-placing 2D materials (and 2D heterostructures) from a source substrate to a target destination has potential applications beyond the creation of pristine designer heterostructures. As advanced functionalities are realized in 2D-based systems, demand will increase for their transfer into other platforms such as nanophotonic systems[152] and photonic integrated circuits [153,154]. Further, advanced and automated characterization can identify specific regions of 2D materials and heterostructures with desired functionalities/properties that can be lithographically isolated, extracted, and then assembled into chip-scale arrays of functional components[146,147,155,156]. Robotic systems based on the advances in heterostructure assembly presented here will be crucial for making immediate steps towards scaling to industrially relevant levels[147]. These advanced manufacturing concepts, combined with the potential scientific discoveries lurking in increasingly sophisticated 2D heterostructures, motivate the continued development of robotic systems for the manipulation, transfer, and stacking of 2D materials.

## 6. Assembling 2D heterostructures in inert atmospheres

Exploration of 2D heterostructures that integrate layers of air-sensitive materials is an important part of 2D materials research. Minimizing oxygen exposure to 2D materials like black phosphorus[157], $FePS_3$[158], $NbSe_2$[6,159], and CVD-grown TMDs[160-162] is critical for the fabrication of high-quality 2D-optoelectronic, superconducting, and magnetic devices/samples. Encapsulating air-sensitive monolayers in polymers[161] and hBN[157] significantly reduces device degradation after fabrication. However, to further improve 2D-heterostructure fabrication, layers should be stacked in an environment free of contaminants like water, oxygen, and hydrocarbons. The ideal solution is to perform all device preparation and characterization in high vacuum (HV). The details and intricacies of stacking and exfoliating in UHV and HV environments are discussed in other sections section. While UHV/HV systems are ideal solutions, the added complexity can be prohibitive. As an intermediate solution, significant improvements to heterostructure fabrication can be achieved by constructing devices in an inert atmosphere of $N_2$ or Ar.

*6.1 – Systems and solutions for assembling heterostructures in inert atmospheres*

The simplest (and cost-effective) approach that has been reported for heterostructure assembly in inert atmospheres used an Ar shielding gas (Figure 8a)[163] while stacking in ambient atmosphere conditions, similar to techniques used to protect fresh welds as they cool in industrial engineering settings. The resulting hBN-encapsulated graphene heterostructures fabricated under these conditions were shown to have fewer bubbles/blisters and subsequently higher carrier mobilities. Further atmospheric control requires the use of a glovebox system (Figures 8b and 8c) such as the early implementation of a "cleanroom in a glovebox"[164]. The use of such larger glovebox "trains" has yielded substantial improvements in heterostructure fabrication. By utilizing separate gloveboxes for lithography, preparation, and characterization[164], cleanroom-quality samples can be fabricated at a fraction of the cost. Nitrogen working environments further reduce costs while maintaining low humidity and oxygen concentrations of less than 20 ppm[165]. Such atmospheric control from fabrication to characterization has been shown to significantly improve the quality of measurements/samples (Figure 8d). However, these systems are cumbersome for directly interacting with instruments. Gloveless anaerobic chambers can be employed to work around the drawbacks of conventional thick-gloved boxes[166]. By utilizing sealed sleeves (Figure 8e), these types of boxes provide users with more dexterity for manual procedures while maintaining oxygen concentrations in the 10s of ppm.

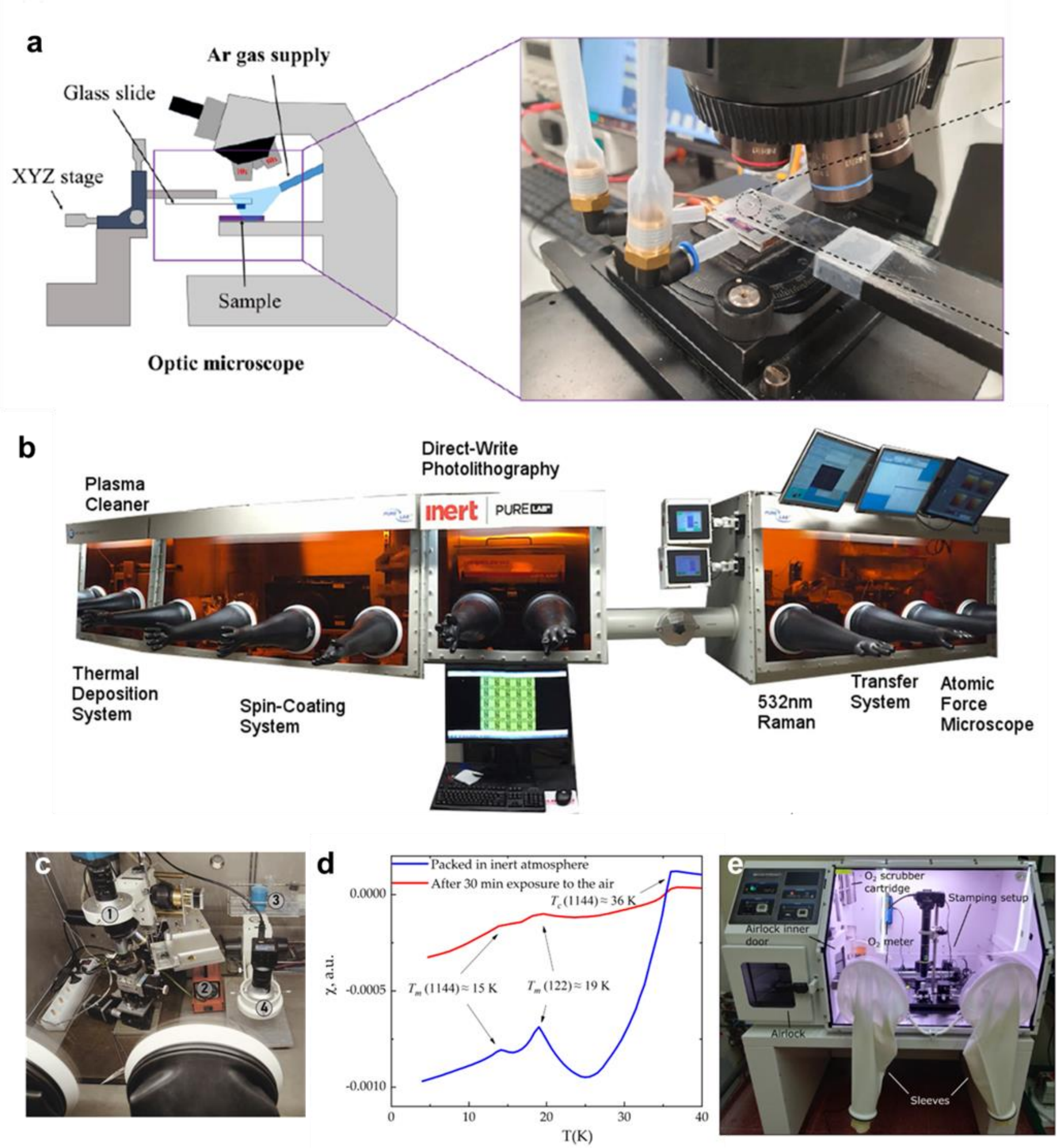

**Figure 8 – Heterostructure assembly under inert atmosphere.** (a) Inert gas can be flowed over the sample and stamp during the assembly process. This approach is relatively fast and yielded high-quality encapsulated graphene devices. (b,c) Heterostructure assembly and device fabrication are adapatable to containment within a glovebox of inert atmosphere, which is a common approach for the fabrication of high-quality devices and processing of air-sensitive materials. (d) Samples can be transferred from the glovebox to cryogenic characterization facilities while minimizing exposure to ambient atmosphere to enable higher-quality meaurements as demonstrated by magnetic susceptibility measurements of $EuCsFe_4As_4$ conducted by Duleba *et al.* [165] (e) Glovebox fitted with an airtight sleeve system that enables the operator to directly use their hand as opposed to the conventional thick-walled glovebox gloves. (a) Reproduced from ref. [163] with permission. (b) Reproduced from ref. [164] with permission. (c) and (d) Reproduced from ref. [165] with permission. (e) Reproduced from ref. [166] with permission.

Finally, in many instances, measurement facilities are not directly connected to the inert atmosphere chambers. Thus, air-sensitive materials require a vacuum or inert atmosphere suitcase for transport to external (out of the glovebox) characterization and modification. These suitcases, often of custom design, can be awkward and cumbersome to maneuver, owing to the arrangement of flanges and chambers required to interface with a particular instrument. For optical characterization, small aluminum two-piece cylinders can be assembled inside the glovebox[167]. The sealed 2D material and assembly can then be characterized with available Raman and PL microscopes outside of the glovebox.

*6.2 – Compatibility of processing components with inert atmosphere*

The major drawback of in-glovebox heterostructure fabrication is the limited compatibility of polymers and solvents with the inert glovebox atmosphere. Solvents like IPA and acetone[168] used in the fabrication of polymer stamps put significant strain on moisture and oxygen absorbers. Sulfuric compounds can even permanently damage copper-based oxygen absorbers. Pick and flip methods using acrylic resins[148] and PVC/PDMS stamps[168] result in an all-dry transfer process that applies to glovebox heterostructure fabrication. Another approach is to use Gel-pak stamps treated with oxygen plasma or UV-ozone outside of the glovebox[169]. These stamps do not result in a temperature-dependent transfer yield, making them excellent for temperature-sensitive 2D materials. A dry transfer technique developed for inert-atmosphere and vacuum heterostructure fabrication utilizes $SiN_X$ membranes[57]. For heterostructures using mechanically exfoliated crystals, these membranes are shaped with optical lithography and reactive ion etching to form protruding cantilevers. The use of other inorganic stamps for inert atmosphere heterostructure stacking is discussed in another section.

To work reliably with air-sensitive materials, exfoliation, inspection, and stacking need to take place entirely in controlled environments such as gloveboxes or vacuum chambers. Robotic tools can perform these tasks with high precision while keeping materials protected from air exposure. Therefore, building modular, glovebox-compatible transfer systems is a practical route toward clean, contamination-free automated assembly, especially when working in UHV/HV environments proves too cumbersome.

### *6.3 – Outlook for inert atmosphere stacking*

Assembling heterostructures inside an inert environment, such as a glovebox, offers a practical alternative to full UHV systems when working with air-sensitive 2D materials. Although UHV platforms can provide the cleanest possible interfaces, glovebox-based setups are more accessible and can support a wider variety of material combinations. However, manual operations inside a glovebox are still slow and depend heavily on the skill of the operator. Introducing robotic manipulation and automated alignment within these inert environments can help reduce variability, limit contamination from handling, and improve overall reproducibility. Building modular transfer and stacking systems that integrate directly with gloveboxes is a realistic and scalable route toward producing high-quality heterostructures without the need for full UHV infrastructure.

## **7. Assembling 2D heterostructures under vacuum**

The sensitivity of many 2D materials to ambient conditions presents a major challenge in achieving high-quality heterostructures[170]. Exposure to oxygen, moisture, and hydrocarbons can degrade 2D materials, introducing defects and trapping contaminants at interfaces. This sensitivity is particularly problematic during the stacking process, where trapped bubbles and residues compromise interfacial quality and alter the material properties. Although inert atmosphere gloveboxes have been widely used to mitigate these effects (see section above), even sub-0.1 ppm contaminants in such environments can negatively impact 2D material properties and interfacial integrity[82,157]. To overcome these limitations, UHV techniques for exfoliation and stacking have emerged as the optimal approach for fabricating pristine 2D heterostructures[33,57,146,157,170-173]. This process requires the development of vacuum-compatible hardware for the different process stages, including exfoliation, flake identification, stacking, and stamp preparation. One can use different combinations of ambient and UHV environments for exfoliation and stacking of 2D materials, or both, as discussed below.

*7.1 – Vacuum Exfoliation*

Exfoliation can be performed in two configurations. One approach is to exfoliate the crystal in ambient conditions using residue-free clean-room adhesive tape[174]. This tape with crystals is then introduced into the vacuum chamber and rolled/pressed over the substrate (Figure 9a), transferring thin flakes onto the substrate (Figure 9b). While the transfer is done under vacuum, this approach, however, still exposes the exfoliated crystals to ambient conditions during the preparation of the tape.

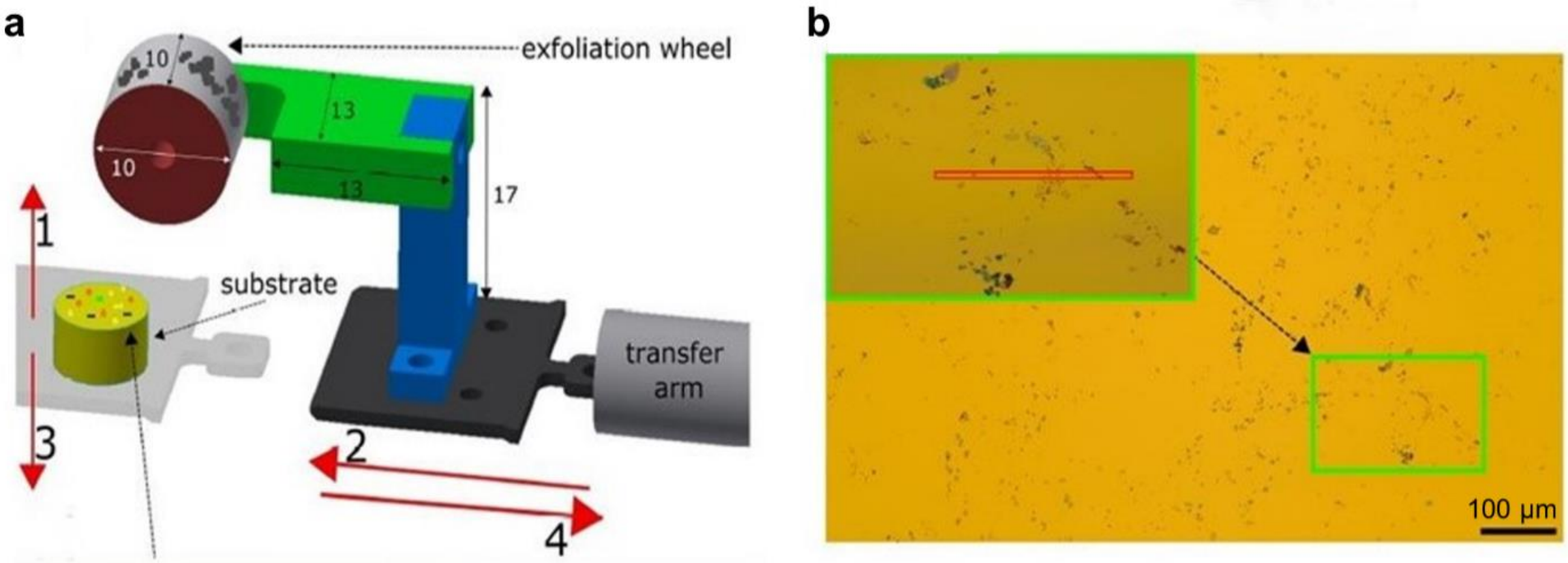

**Figure 9 – In-situ exfoliation.** (a) schematic of the in-situ exfoliation set-up. (b) optical image of thin $MoS_2$ exfoliated on substrate after three exfoliation cycles. Scale bar 100 µm. (a) and (b) reproduced from ref. [170] with permission.

Alternatively, to eliminate exposure to ambient conditions, exfoliation can be performed directly in HV or UHV. This approach requires the bulk crystal to be cleaved under vacuum[81,82]. Substrate preparation involves sputtering, annealing, and plasma cleaning (Figure 10, step 1), and depending on the material being exfoliated, the substrate material can be varied (e.g., Si, Au, etc). To perform the exfoliation, the freshly cleaved crystal and the substrate are pressed together and then separated (Figure 10, Steps 2 and 3). This process has been used to exfoliate several 2D materials ($WSe_2$, $WS_2$, $WTe_2$, and AgTe) under vacuum conditions. Crystal preparation under vacuum not only produces a high-quality film but also enhances the size of exfoliated atomically thin films up to millimeters[171].

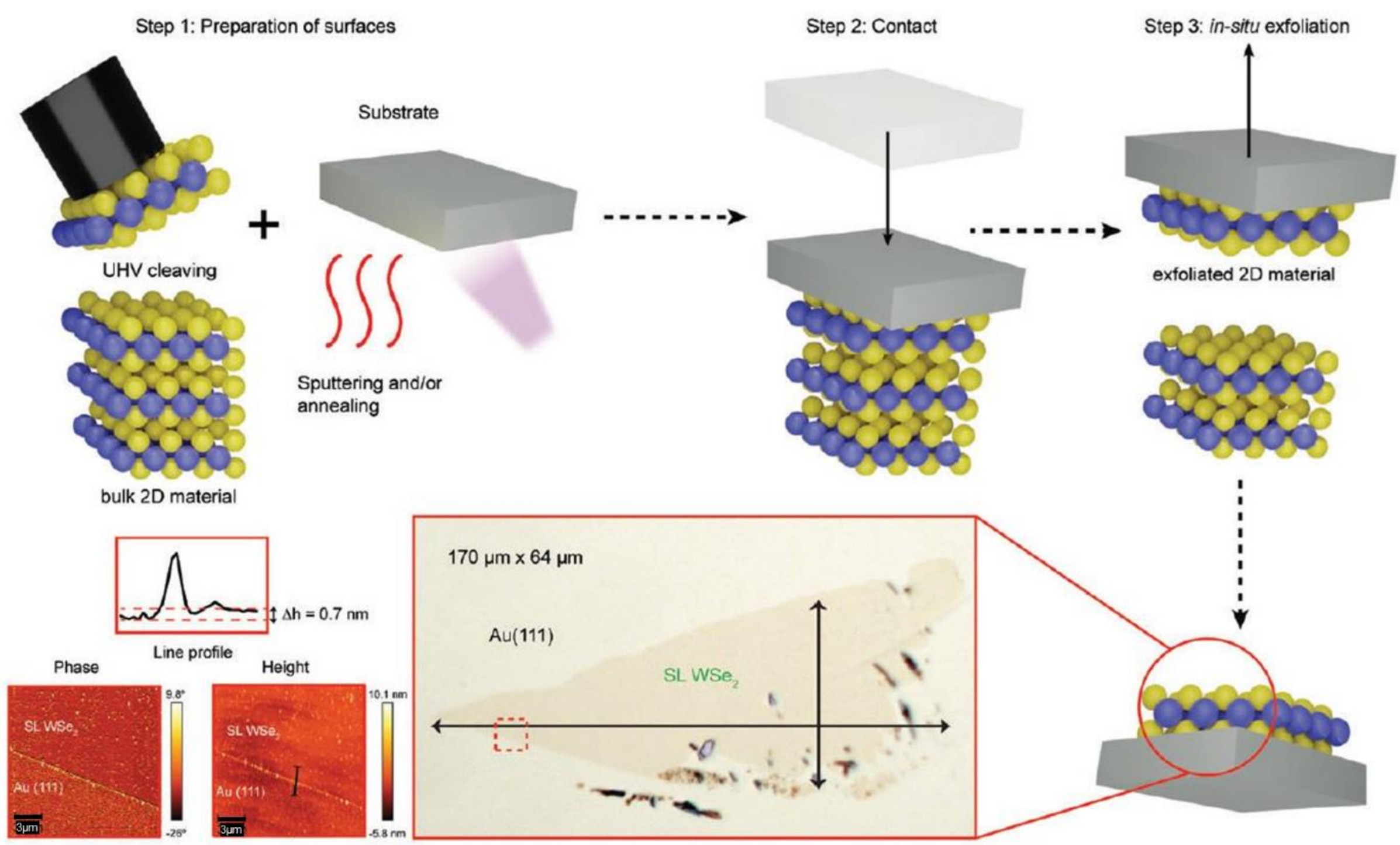

**Figure 10 - Schematic of the in-situ exfoliation and transfer of 2D materials.** Step 1 demonstrates a sample surface cleaved in UHV to expose freshly exfoliated crystal. The substrate is sputtered and/or annealed to create a clean and flat surface. Step 2 demonstrates crystal and substrate brought in contact by pressing them together. In step 3, the crystal and substrate are separated resulting in exfoliation of 2D material on substrate. Optical microscope displays a large area of monolayer $WSe_2$ on top of the substrate. AFM scans confirm the characteristic height on monolayer $WSe_2$ of 0.7nm. This panel reproduced from ref. [81] with permission.

*7.2 – Vacuum Stacking*

Two vacuum-based systems have been developed for stacking 2D materials. Imamura *et al*.[172] created a HV system with a pressure of $10^{-6}$ mb, where two layers are pressed together at 200°C for one hour, resulting in high-quality transfer verified by optical microscope, μ-Raman spectroscopy, LEED, and ARPES. However, this method has a drawback—it is difficult to control the twist angle between the layers, and precise stacking is not possible. Mannix *et al.*[146] designed a more advanced robotic system that operates in HV ($10^{-6}$ Torr) (Figure 7d) and can assemble heterostructures with up to 80 layers. This system uses a multilayer polymer stamp (Figure 11a), which heats and cools through specific temperature changes to pick up and drop each layer (Figure 11b). The stamp consists of PDMS, a support layer lift-off resist, and a release layer made of poly(cyclohexene propylene carbonate) (PCPC), along with an adhesion layer of poly(benzyl methacrylate) (PBzMA). While this system has been shown to work with different materials like CVD-grown TMDs (e.g., 16 layers of $MoS_2$ shown in Figure 11c and Figure 11d), exfoliated graphene, and boron nitride, its main limitation is the need for polymer cleaning with acetone or chloroform after stacking, which is not vacuum compatible.

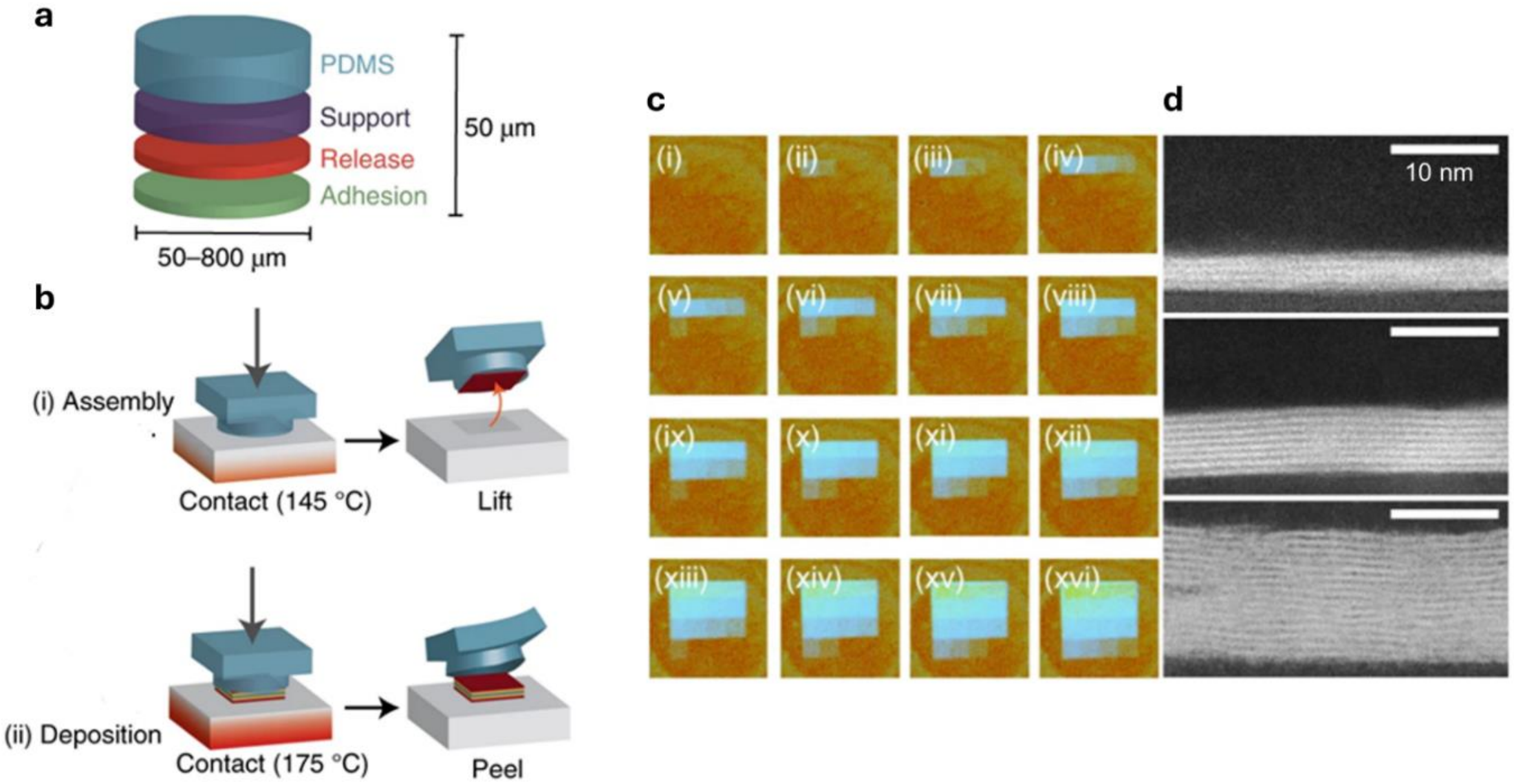

**Figure 11 - Pixel-based stamping in vacuum.** (a) Different layers of stamps used for assembly and deposition of 2D crystals b) Schematic demonstrating different steps for the assembly (i) and deposition (ii) of 2D heterostructures on a target substrate c) Optical images of the 16-tile heterostructure assembly. Each image displays a layer (i) through (xvi). The images were taken when the stamp was in contact with the substrate d) Cross-sectional STEM images of 4 L (top), 8 L (middle) and 16 L (bottom) $MoS_2$ layers demonstrating atomically resolved thickness control. All scale bars, 10 nm. (a)-(d) reproduced from ref. [146] with permission.

*7.3 – Exfoliation and stacking in an ultrahigh vacuum environment*

UHV is the ultimate environment to eliminate the exposure of 2D flakes to contaminants during exfoliation and stacking. Guo *et al.*[171] developed a UHV system ($5 \times 10^{-10}$ mb) that included both exfoliation and stacking under UHV. This system was directly connected to an MBE growth chamber. The system consists of two main stages: a tool stage and a sample stage (Figure 12a). Samples are placed on the sample stage, picked up by the tool stage, and precisely positioned onto the substrate, which is also on the sample stage. The entire assembly process is optically accessible through a long working distance objective. Exfoliation is carried out using UHV-compatible Kapton tape onto a silicon oxide or gold-coated silicon substrate (Figure 12b). A stamp made of sapphire, PDMS, PMMA, and polypropylene carbonate (PPC) is pre-assembled and degassed before being introduced into the UHV chamber. This stamp is then used to pick up exfoliated monolayers and stack them into heterostructures (Figure 12c). The system has demonstrated BP/hBN heterostructures with all operations occurring under vacuum (Figure 12d). However, the use of polymers still results in some residual contamination, visible as bright specks in AFM images. To circumvent the limitation in polymer-based stamps, Wang *et al.* developed and demonstrated a UHV transfer using inorganic stamps. The details of the stamp and the key layers for picking up and transferring thin layers of 2D materials are described in the next section.

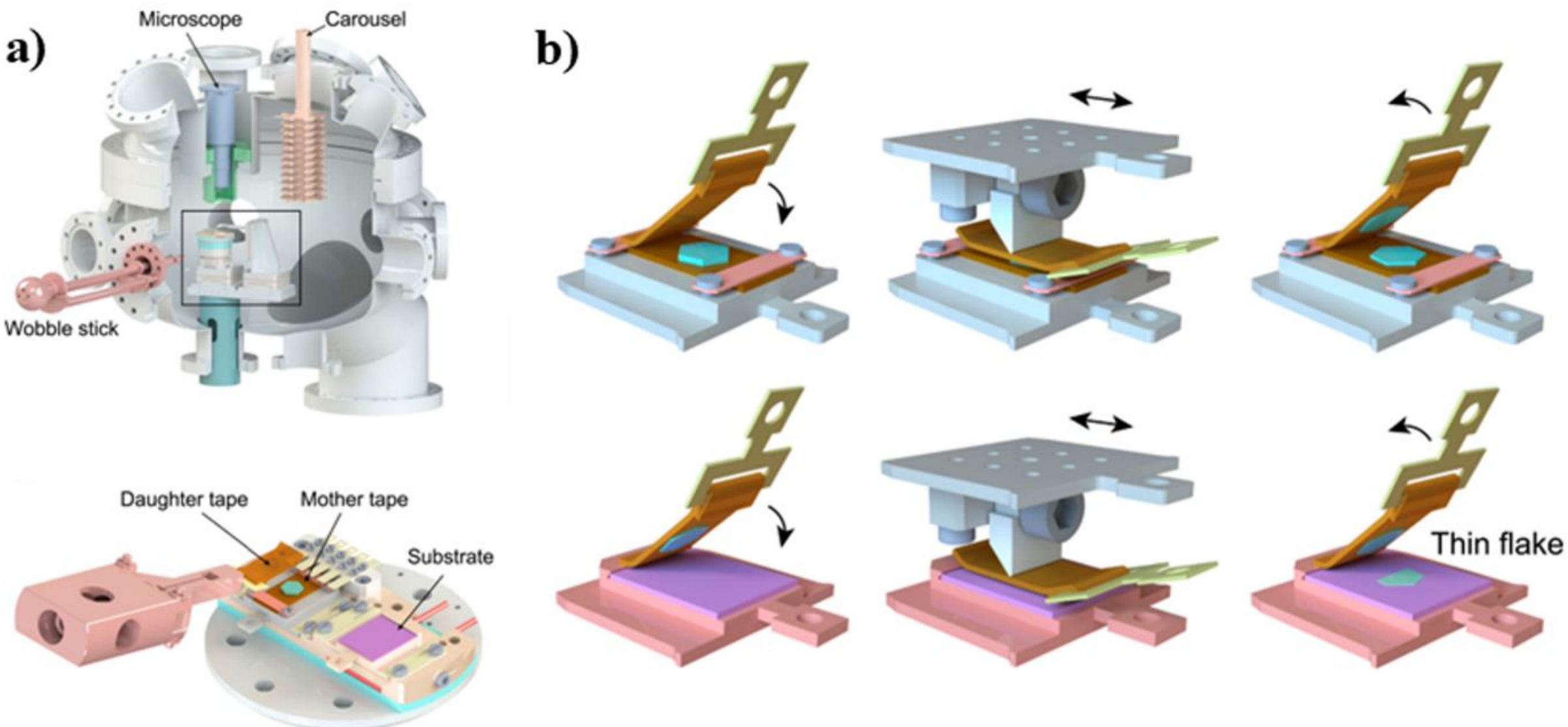

**Figure 12 – Mechanized exfoliation and stacking in UHV.** (a) Top: Design of the fabrication setup. Set-up has two stages: a sample stage and a tool stage. A microscope (blue) mounted outside of the chamber is used to monitor the motion of the sample and the tool through the re-entrant viewport (green). The fabrication setup is equipped with a 72-slot carousel for sample/tool storage (gold). A wobble stick (brown) is used to shuffle sample/tools among the sample and tool stages, and the carousel and the transfer rod from the MBE chamber. Bottom: Schematic illustration of the setup during exfoliation of 2D materials. The mother and daughter tapes carrying pieces of layered crystals are prepared on the left slot. The substrate is held on the right slot where the exfoliation is performed. (b) Step-by-step illustration of a complete 2D material exfoliation process. Top row three configurations: A daughter tape carrying pieces of layered crystal is prepared by pressing the tape against a mother tape with a wedge. Bottom three configurations: show the daughter tape is pressed against the substrate and then lifted up to exfoliate 2D material flakes on the substrate. (a) and (b) reproduced from ref. [171]with permission.

*7.4- Outlook*

UHV and HV systems allow for extremely clean interfaces and precise control during stacking, which makes them one of the most dependable methods for producing high-purity van der Waals heterostructures. However, current setups are still slow, require careful manual alignment, and are not designed for high-throughput production. Adding robotic alignment, controlled contact forces, and real-time monitoring could significantly improve consistency while lowering dependence on operator skill. As these capabilities develop and modular UHV cluster tools become easier to access, UHV/HV assembly could shift from a specialized laboratory technique to a scalable platform for automated device manufacturing.

## 8. 2D material transfer and heterostructure assembly with inorganic stamps

As discussed in the introductory sections, exfoliation is commonly performed with organic adhesives, and dry transfer stamps are most commonly composed of organic materials such as PMMA, PVA, PC, or PDMS[33]. If an intermediate layer of a heterostructure is directly contacted by any of these organic materials, the corresponding interface will be heavily contaminated with residue and bubbles/blisters. Currently, the primary approach to minimize organic contamination is the van der Waals pickup method (discussed in detail in the introduction) where the heterostructure is assembled on an organic stamp. Here, hBN is typically the first layer and can be thought of as an inorganic stamp: although organic materials are used to pick up the hBN, all of the subsequent layers use the hBN crystallite to avoid direct contact and contamination with organic materials[2]. However, for surface-sensitive measurements, this approach will leave the top hBN surface of the heterostructure contaminated with polymer materials. Additionally, making electrical contact to the layers of the heterostructure is a challenge since they are encapsulated by insulating h-BN layers. To mitigate these issues, as well as improve the yield of exfoliation processes, there have been a host of recent efforts to replace organic materials with inorganic materials to exfoliate and pick up 2D materials. As these processes mature there will be an increasing demand for their integration into automated systems and workflows for 2D materials research.

*8.1 – Pure metal-assisted stamping and exfoliation*

Exfoliation and stamping/transfer processes that rely on continuous metal films are well-established organic-free alternatives for sample preparation and heterostructure assembly. Empirically, many studies have demonstrated that the adhesion between single layers of 2D materials and metal surfaces can be significantly stronger than the interlayer adhesion in a bulk crystal (and the adhesion of a CVD-grown single-layer to a growth substrate). Therefore, when a pristine metal film is brought into contact with a bulk crystal of 2D materials or a large-area continuous film, it can efficiently extract single layers from the source. The extracted single-layer crystallites can reach macroscopic dimensions, bridging laboratory-scale mechanical exfoliation to large-area production for intermediate-scale manufacturing of 2D materials derived from high-quality bulk crystals. Many review articles on metal-assisted exfoliation [175-177] and metal-assisted transfer [26,31,138] have been recently published, providing comprehensive portraits of the state-of-the-art for human-powered metal-assisted exfoliation and transfer techniques. Further, significant effort has been invested into establishing a microscopic understanding of the 2D material-metal interactions and is covered in a recent review by Pirker *et al.* [178]. Here, we focus specifically on adapting these techniques to prepare crystallites/heterostructures with high-quality surfaces/interfaces and on their implementations into automated systems.

Metal-assisted exfoliation/transfer to achieve a *single* ultraclean interface with lateral dimensions as large as the bulk crystal (or continuous CVD-grown film) is now relatively straightforward. In general, processes follow a similar protocol to Kim *et al.* [179], where Ni was deposited on (predominantly) single-layer graphene grown on a SiC substrate. The Ni/graphene was then exfoliated from the SiC substrate using thermal release tape, exposing a graphene surface that was kept pristine by the underlying SiC substrate. For TMD materials, the use of Au in a similar fashion enables exfoliation of large-area single-layer crystallites from bulk crystals [180]. While Au is the most common metal used, other metals such as Ag and Ni have been shown to also work under the right conditions [181-183]. Very recently, Shen *et al.* adapted the Au-assisted procedure to prepare a surface of CVD-grown 1L-$MoS_2$ that was sufficiently pristine for high-quality XPS, STM, and ARPES measurements [184]. Outside of the UHV environment needed for surface measurements, Au was deposited on as-grown 1L-$MoS_2$ on sapphire, but rather than adhering the top surface of the Au film to epoxy or thermal release tape, the sapphire/1L-$MoS_2$/Au substrate was glued directly to a sample holder. The sample holder was then transferred into a UHV chamber where a mechanized rod was used to remotely "exfoliate" the sapphire growth substrate from the 1L-$MoS_2$/Au system. The high quality of the sample—namely the pristineness of the exposed surface of the 1L-$MoS_2$—in combination with surface-sensitive measurements unveiled a new charge density wave in $MoS_2$, highlighting the utility of the metal-assisted transfer/exfoliation techniques. While clearly valuable, automation of these processes in the near term will be challenging because the robotic tools discussed above will need to be tightly integrated with metal deposition systems, adding another significant requirement/barrier to the development task.

From the perspective of near-term automation, it is more simple to utilize a metallic film in place of the adhesive tapes based on organic materials. Such a metallic tape could be pre-manufactured and more readily used in robotic exfoliation systems. Several groups have reported exfoliation of macroscopic single-layer crystallites of TMDs using a gold film in place of adhesive tape [64,185-187]. To be successful, however, the adhesion between the top layer of the 2D crystal and the gold needs to be maximized by minimizing surface roughness and contamination. For instance, Velicky *et al.* used smooth sputtered Au films to exfoliate a large catalog of 2D TMDs but observed that exposure of the gold film to ambient atmosphere for more than 6 minutes rendered the procedure ineffective [186]. Such temporal urgency complicates automation. Alternatively, both Magda *et al.* [187] and Liu *et al.* [185] deposited a gold film onto a template substrate (mica and polished $SiO_2$/Si, respectively), importantly, without an adhesion layer. Before exfoliation, the gold film was stripped from its substrate, exposing a pristine surface with a roughness that mirrors the template. Immediately after, bulk crystals were adhered to the gold surface, optionally annealed, and then removed using thermal release tape. Between the two groups, large-area exfoliation of several TMDs onto a gold substrate was demonstrated. Such template-stripped metal films are more amenable to automation: the films can be prepared beforehand in mass quantities, and adding template-stripping immediately before exfoliation/transfer is a reasonable expansion of existing automated workflows. Taking the first steps in this direction, Galafassi *et al.* recently reported a process where tailored metal stamps are stripped from their template substrate using a PDMS stamp [188], laying important groundwork for future automation.

When it comes to the assembly of heterostructures, however, a major challenge with metal-assisted exfoliation/transfer is that the strong binding between the metal and TMDs complicates the removal of the metallic film and typically requires a wet-chemical etching step. Consequently, the 2D materials are exposed to solvents, which can contaminate their surfaces, and if the etching is incomplete, leave remnant metal contamination/doping. Seemingly overcoming this challenge, Zaborski *et al.*[189] recently reported a fabrication procedure that uses both Au-assisted exfoliation and transfer to produce high-quality, large-area twisted bilayers of $MoS_2$ with astonishing dimensions of ~2×3 $mm^2$. The homogeneity of the twist angle between the two layers was confirmed across the sample using point STM and piezo force microscopy characterization, and the emergence of minibands (signatures of strong interlayer coupling) was demonstrated using ARPES. Combined, these measurements confirm the assembly of a large-area, high-quality twisted heterostructure. Their process consists of multiple steps, three of which are chemical etching but through careful design, the 2D materials are always protected from direct exposure to organic solvents or polymer materials by the Au layers. Removal of the Au layers requires that 2D materials are exposed to Au etchant, in which the authors claim that $MoS_2$ is stable. This stability enables the etching process to run to completion, and if the deposited gold is sufficiently pure (a non-trivial challenge in a typical shared-use facility), the 2D materials will be left uncontaminated by any residual metals. While the large area heterostructures produced are very appealing, automation will be challenging because of the multiple wet chemical etching steps and the need for two types of stamps. In addition, the process is only able to generate a heterostructure with one pristine interface, limiting its current form for the assembly of more complicated 2D systems. This latter challenge, however, may be on the precipice of being solved: recent efforts to clean these exposed surfaces have shown promising results for expanding the process to the assembly of additional layers and interfaces [190].

*8.2 – Hybrid metal stamps for 2D transfer and heterostructure assembly*

Whereas the preceding section focused on using continuous metal films to facilitate the exfoliation of large areas of single-layer 2D materials, the enhanced adhesion of the metals to TMDs (and presumably other 2D materials) can be more precisely leveraged to improve the assembly of 2D heterostructures by optimizing the structure of the stamp. The common theme is to tune the thickness of the metal film to both optimize the adhesion force and insulate the 2D materials from direct contact with any polymers or solvents used. Because these approaches focus on engineering the stamp, they can, in general, be integrated into robotic stacking systems with relative ease. Here, we highlight a few recent reports that could be the start of new generations of stamp structures worth considering in future automation efforts.

Recently, Kim *et al.* synergized the van der Waals pickup method with gold-assisted transfer to fabricate a twisted trilayer graphene (TTG) heterostructure with a top electrode. In many ways, this structure is a holy grail: it demonstrated strong interfacial coupling, had a top gate, and had a pristine top surface for scanning tunneling microscopy (STM). This process was successful because of the weak interaction between Au and graphene (Figures 13a-f)[191]. In their work, a stack of graphite/hBN/TTG was assembled on a PDMS/PC stamp using the van der Waals pickup method. However, to investigate the properties of TTG by STM, the entire structure needed to be flipped to make the TTG surface accessible to the STM probe. To obtain the flipped sample, the graphite/hBN/TTG was transferred to a PDMS/Au substrate (Figure 13c). The thickness of the gold was only 12 nm (plus an additional 3 nm titanium adhesion layer), which proved sufficient to protect the TTG surface from the PDMS and solvents used to remove the PC but was only weakly adhesive to the TTG. After removing the PC, the graphite/hBN/TTG/Au/PDMS was transferred to a silica substrate. The interaction between graphite and silica is stronger than that between TTG and Au, so the PDMS/Au capping layer was easily removed (Figure 13f), leaving a TTG heterostructure with a pristine surface for STM measurements. To attach a top electrode to the TTG (for back-gating), the group also used PDMS to stamp a continuous gold film onto a portion of the TTG without contaminating the surface with polymers. Such an inversion process and electrode stamping should be relatively easy extensions to existing automated stacking workflows.

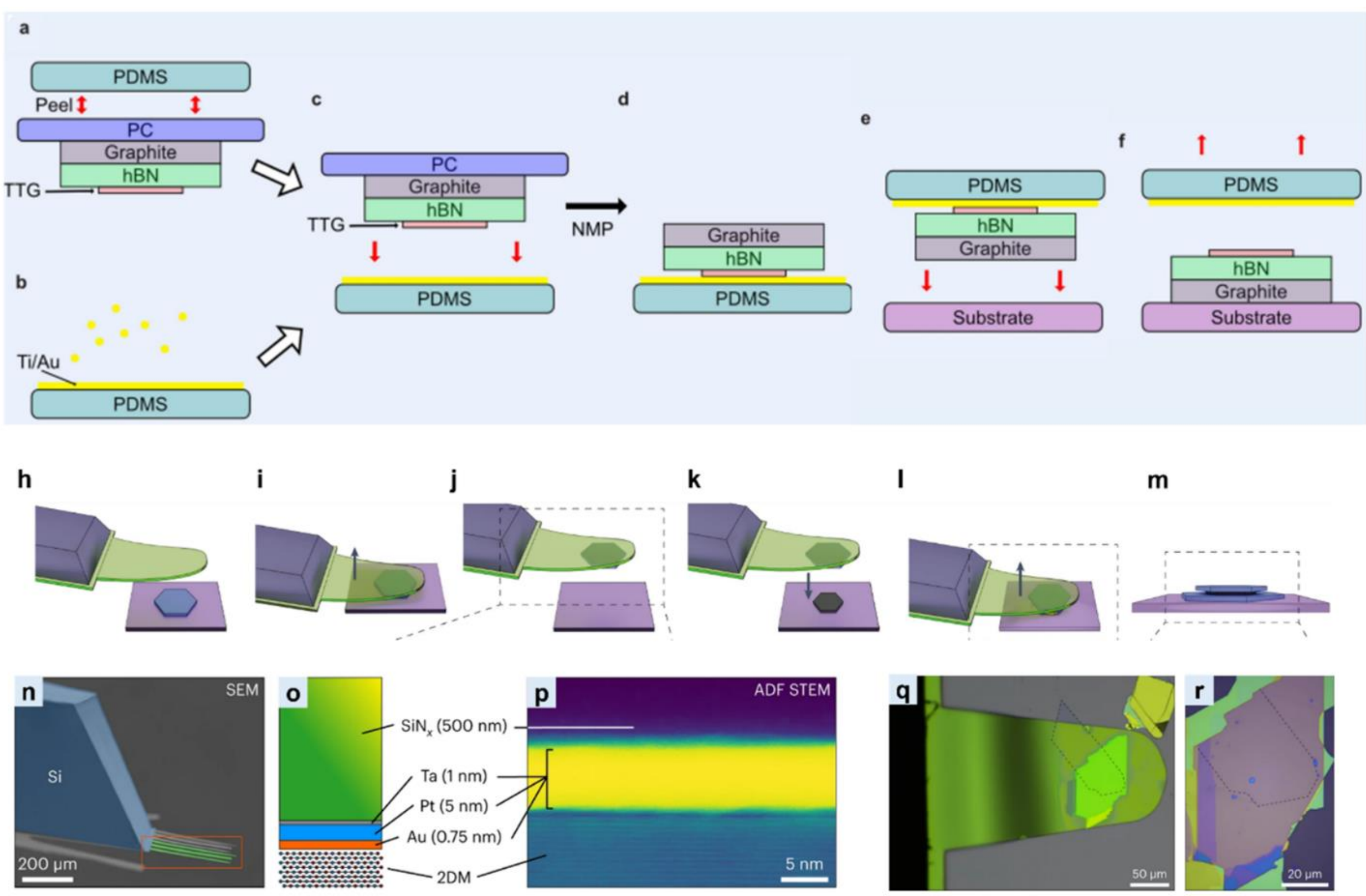

**Figure 13 - Heterostructures by inorganic stamping using an Au thin layer.** (a)-(f) Au-assisted flipping twisted trilayer graphene (TTG) for gate-tunable STM measurement. (h)-(r) Au thin film on $SiN_x$ membrane for a general approach to fabricate 2D heterostructures. (a)-(f) reproduced from ref. [191] with permission. (h)-(r) reproduced from ref. [57] with permission.

A similar approach to eliminating organic material was recently reported by Wang *et al.*[57]. In their work, the adhesion between the Au film and 2D materials is tuned by the thickness of the Au film so that the adhesion is sufficiently strong to pick up 2D flakes from a substrate, but weaker than interlayer adhesions so that flakes can be transferred to the heterostructure. In their case, the optimum thickness of gold is ~0.65 nm. The steps of the process are summarized in Figures 13h-13m. A SiN membrane attached to a cantilever (Figure 13n) was used to precisely control the contact between the Au layer and 2D flakes.

Using this technique, the samples are almost free of bubbles, which is difficult to obtain when the process involves organic materials, marking an important achievement for the fabrication of large-area uniform 2D heterostructures (Figures 13o-r). One challenge with the SiN-based stamps is that the thickness of the Au film that is responsible for the adhesion needs to be optimized for each material. However, the quality of the heterostructures produced seems to wholly justify the effort and investment in the optimization process.

### *8.3 – Outlook for organic-free exfoliation and heterostructure assembly*

The contamination from organic stamping and exfoliation materials is one of the most ubiquitous challenges to fabricating high-quality heterostructures as well as the preparation of single-layer samples that are suitable for surface-sensitive measurements such as AFM, STM, and ARPES. The recent advances in metal-assisted exfoliation/transfer and organic-free stamping procedures are encouraging prospects for a future where organic materials are completely removed from the preparation and heterostructure fabrication processes. However, significant work is still needed to come up with organic-free processes that are as versatile as their counterparts that are based on organics. For instance, the Au-assisted fabrication processes are immensely appealing to integrate with automated 2D heterostructure fabrication. However, it will be a significant challenge to automate these procedures such that the "freshness" of the metal surfaces is preserved. We note that a recent study reported that oxygen plasma cleaning renews the cleanliness of the gold film [192]. Further, the metal-assisted techniques are not universally applicable to all 2D materials, so a system that implements them will not be as versatile as relying on adhesive tapes/films. Nevertheless, tackling this challenge is worth the substantial investment of effort. If successful, synergizing metal-assisted exfoliation and transfer methods with automation promises a pathway toward consistent, high-throughput manufacturing of clean, functional heterostructures that could be suitable feedstock for scalable manufacturing processes that involve 2D crystallites and heterostructures.

## 9. Toward complete automation

Achieving full automation in 2D crystallite preparation and heterostructure assembly requires linking exfoliation, flake identification, transfer, and stacking into one coordinated workflow rather than treating them as separate steps. Table 1 summarizes the semi-automated systems that have been reported to date and discussed in this review. Although automated exfoliation systems, machine-learning flake detection, and robotic transfer platforms already exist, they are still predominantly used and developed as stand-alone modules to tackle one step of the process. Notable exceptions include the 2DMMS[60] system developed by Masubuchi *et al.* and the QPress system[119] that is operated by Brookhaven National Lab. However, QPress requires human intervention to, for instance, exchange samples between the instruments, and the 2DMMS does not include an automation solution for exfoliation. While both systems are laudable advances for 2D materials preparation, their respective automation gaps need to be overcome in the next generations of the instruments. To that end, groups working on robotic instruments should increasingly prioritize the "automation readiness" of their systems, which includes their programmability through application programming interfaces (APIs), their compatibility with robotic arms to facilitate sample exchange, and their ability to autonomously execute processes for hundreds of samples.

**Table 1.** Comparative overview of exfoliation and stacking methods used for 2D material heterostructure fabrication. The table summarizes each approach in terms of advantages, limitations, reproducibility, scalability, and automation readiness.

| **Method and description** | | | **Comparative ranking of process metrics** | | | |
|---|---|---|---|---|---|---|
| **Method** | **Processes** | **Op. Env(s).** | **Primary advantages**<br>**Process limitations** | **Reprod.** | **Scal.** | **Auto. Read.** |
| Manual | Exfoliation<br>Cataloging<br>Stacking | Ambient<br>Glovebox | Simple, inexpensive<br>Operator-dependent, significant variability, time-consuming | Low | Low | None |
| Rheometer-based copy & paste [66] | Exfoliation | Ambient | Tunable torque and pressure for copy-and-paste operation<br>Requires manual transfer/exfoliation of material to substrate | Low-Mod. | Low | Partial |
| eXfoliator [68] | Exfoliation | Ambient | Precisely tunable peeling rate and temperature<br>Fixed exfoliation motion, incomplete process optimization | Mod. | Mod. | Low |
| Gear-and-rack exfoliator [67] | Exfoliation | Ambient | Adjustable peel geometry improves yield<br>Narrow material set demonstrated, incomplete process automation | High | Mod. | Low |
| Reel-to-reel exfoliator [70] | Exfoliation | Ambient | High-throughput<br>Unsuitable for producing single-layer crystallites | Low | High | High |
| Liquid-phase exfoliation [83] | Exfoliation | Liquid | Scalable production based on solution processing<br>Unsuitable for producing single-layer crystallites | Low | High | Limited |
| 2DMMS [60] | Cataloging | Glovebox | Semi-autonomous robotic pick-and-place | High | High | Partial |

| Method and description | | | Comparative ranking of process metrics | | | |
|---|---|---|---|---|---|---|
| **Method** | **Processes** | **Op. Env(s).** | **Primary advantages**<br>**Process limitations** | **Reprod.** | **Scal.** | **Auto. Read.** |
| | Stacking | | Relies on manual exfoliation, operator supervision is required | | | |
| QPress [119] | Exfoliation<br>Cataloging<br>Stacking | Glovebox | Integration of all processes, demonstration of high-quality heterostructures<br>Limited readiness for autonomous operation, incomplete process optimization | High | Mod | Low |
| VAR [146] | Stacking | Vacuum | Automated assembly of $N$-layer heterostructures in vacuum<br>Designed for CVD-grown materials, limiting versatility for the entire 2D materials library, sample exchange with vacuum may be challenging | High | Mod | Low |

Op. Env(s): Operating environment(s)
Reprod: Process reproducibility for 2D heterostructure fabrication
Scal: Process scalability
Auto. Read.: Process automation readiness

Further progress toward complete automation depends on real-time sensing of physical properties such as flake thickness, adhesion behavior, and interface cleanliness. Such sensing can be integrated with adaptive feedback that can dynamically adjust mechanical and thermal conditions during pickup and placement. Progression will also require instrumentation development that has the versatility to integrate new processing innovations such as organic-free exfoliation and transfer. This type of versatility can substantially increase the complexity of the instruments, highlighting the need for teams to include experts who are familiar with the latest state-of-the-art in robotics, automation, and, increasingly, AI. Finally, leveraging the full library of 2D materials will require that at least some version of these tools can operate in gloveboxes or UHV cluster systems that prevent oxidation and contamination throughout the entire process, as discussed above. Integrating these capabilities into modular, closed-loop platforms would allow heterostructures to be assembled cleanly and consistently without manual handling, providing a realistic pathway toward scalable and repeatable device manufacturing.

## 10. Conclusions and outlook

This review paints a portrait of a vibrant subgroup of the 2D materials research community that is making substantial advances that will lead to AI-driven automation for the preparation of high-quality, large-area individual 2D systems and the assembly of 2D heterostructures. From uniting these advances into new generations of instruments and processes, we should expect (and demand!) faster fabrication rates, higher-quality heterostructures, community-wide access to more complex/sophisticated many-layer heterostructures, and significant leaps in reproducibility that are enabled by the dramatic improvement of fabrication capabilities. The tools and process innovations presented here, if nurtured and further developed upon, possess the potential to transform experimental research of 2D materials from long odysseys of repeated exfoliation and stacking for undergraduate and graduate students to projects where sample production is as routine and simple as, for instance, the fabrication of nanophotonics and integrated photonic circuits.

Furthermore, the tools highlighted here, especially the robotics and AI algorithms, are a solid foundation on which to develop fully automated, AI-driven, 2D materials research facilities, similar to other emerging automated laboratory projects[193]. In such an automated laboratory, the synthesis of bulk crystals, their processing/exfoliation into single-layer samples, stacking into complex heterostructures, packaging into devices, measurement of physical properties, data analysis, and then hypothesis formulation could ultimately be autonomously conducted for rapid, continuous exploration of the massive parameter space of 2D materials. Self-driving capabilities will be enhanced by the inclusion of models and theoretical analysis of the key physical mechanisms, such as adhesion and fracturing, that underlie the 2D materials preparation workflow[194,195]. These robotic tools could also serve as a bridge between the one-off fabrication processes of today to semi-scalable manufacturing, where arrays of functional 2D material systems can be assembled and then heterogeneously integrated as a single functional component in a larger device architecture. Further, such a self-driving and autonomous facility could be used in a closed-loop form to test theoretical hypotheses and predictions, greatly accelerating the pace of discovery and analysis of new 2D materials, heterostructures, and emergent phenomena.

With such multifaceted potential, the development of these tools should be prioritized. But it should also be recognized that these development projects need truly multidisciplinary teams composed of scientists, engineers (electrical, mechanical, industrial, etc.), and software developers. They are, by definition, ambitious and large projects. Following the precedent of Masubuchi *et al.*[60], the rate of development will be greatly accelerated through the dissemination of software, design files, bills of materials, neural networks, training data, etc. through modern open-source repositories. Finally, community-wide efforts, to, for instance, standardize AI training data, adopt common software platforms, and convene technical working groups and focus sessions at conferences, will also contribute to the acceleration of development and encourage widespread adoption in other groups, which will, in turn, naturally lead to faster community-driven development of these soon-to-be critical tools.

**Acknowledgments**

We acknowledge the MonArk NSF Quantum Foundry supported by the National Science Foundation Q-AMASE-i program under NSF award No. DMR-1906383.

**References:**

1 Dean, C. R. *et al.* Boron nitride substrates for high-quality graphene electronics. *Nature Nanotech.* **5**, 722-726 (2010). https://doi.org/10.1038/nnano.2010.172

2 Wang, L. *et al.* One-dimensional electrical contact to a two-dimensional material. *Science* **342**, 614-617 (2013). https://doi.org/10.1126/science.1244358

3 Huang, B. *et al.* Layer-dependent ferromagnetism in a van der Waals crystal down to the monolayer limit. *Nature* **546**, 270-273 (2017). https://doi.org/10.1038/nature22391

4 Gong, C. *et al.* Discovery of intrinsic ferromagnetism in two-dimensional van der Waals crystals. *Nature* **546**, 265-269 (2017). https://doi.org/10.1038/nature22060

5 Doganov, R. A. *et al.* Transport properties of pristine few-layer black phosphorus by van der Waals passivation in an inert atmosphere. *Nat. Commun.* **6**, 6647 (2015). https://doi.org/10.1038/ncomms7647

6 Cao, Y. *et al.* Quality Heterostructures from Two-Dimensional Crystals Unstable in Air by Their Assembly in Inert Atmosphere. *Nano Lett.* **15**, 4914-4921 (2015). https://doi.org/10.1021/acs.nanolett.5b00648

7 Zibrov, A. A. *et al.* Tunable interacting composite fermion phases in a half-filled bilayer-graphene Landau level. *Nature* **549**, 360-364 (2017). https://doi.org/10.1038/nature23893

8 Cao, Y. *et al.* Unconventional superconductivity in magic-angle graphene superlattices. *Nature* **556**, 43-50 (2018). https://doi.org/10.1038/nature26160

9 Avsar, A. *et al.* Spin-orbit proximity effect in graphene. *Nat. Commun.* **5**, 4875 (2014). https://doi.org/10.1038/ncomms5875

10 Wang, Z., Tang, C., Sachs, R., Barlas, Y. & Shi, J. Proximity-induced ferromagnetism in graphene revealed by the anomalous Hall effect. *Phys. Rev. Lett.* **114**, 016603 (2015). https://doi.org/10.1103/PhysRevLett.114.016603

11 Efetov, D. K. *et al.* Specular interband Andreev reflections at van der Waals interfaces between graphene and $NbSe_2$. *Nat. Phys.* **12**, 328-332 (2015). https://doi.org/10.1038/nphys3583

12 Yasuda, K., Wang, X., Watanabe, K., Taniguchi, T. & Jarillo-Herrero, P. Stacking-engineered ferroelectricity in bilayer boron nitride. *Science* (2021). https://doi.org/10.1126/science.abd3230

13 Cao, Y. *et al.* Correlated insulator behaviour at half-filling in magic-angle graphene superlattices. *Nature* **556**, 80-84 (2018). https://doi.org/10.1038/nature26154

14 Zhao, W. *et al.* Gate-tunable heavy fermions in a moire Kondo lattice. *Nature* **616**, 61-65 (2023). https://doi.org/10.1038/s41586-023-05800-7

15 Padhi, B., Chitra, R. & Phillips, P. W. Generalized Wigner crystallization in moiré materials. *Physical Review B* **103** (2021). https://doi.org/10.1103/PhysRevB.103.125146

16 Tang, Y. *et al.* Evidence of frustrated magnetic interactions in a Wigner-Mott insulator. *Nature Nanotech.* **18**, 233-237 (2023). https://doi.org/10.1038/s41565-022-01309-8

17 Ji, Z. *et al.* Local probe of bulk and edge states in a fractional Chern insulator. *Nature* **635**, 578-583 (2024). https://doi.org/10.1038/s41586-024-08092-7

18 Zeng, Y. *et al.* Thermodynamic evidence of fractional Chern insulator in moire $MoTe_2$. *Nature* **622**, 69-73 (2023). https://doi.org/10.1038/s41586-023-06452-3

19 Park, H. *et al.* Observation of fractionally quantized anomalous Hall effect. *Nature* **622**, 74-79 (2023). https://doi.org/10.1038/s41586-023-06536-0

20 Cai, J. *et al.* Signatures of fractional quantum anomalous Hall states in twisted $MoTe_2$. *Nature* **622**, 63-68 (2023). https://doi.org/10.1038/s41586-023-06289-w

21 Lu, Z. *et al.* Fractional quantum anomalous Hall effect in multilayer graphene. *Nature* **626**, 759-764 (2024). https://doi.org/10.1038/s41586-023-07010-7

22 Kang, K. *et al.* Evidence of the fractional quantum spin Hall effect in moire $MoTe_2$. *Nature* **628**, 522-526 (2024). https://doi.org/10.1038/s41586-024-07214-5

23 Lu, Z. *et al.* Extended quantum anomalous Hall states in graphene/hBN moire superlattices. *Nature* **637**, 1090-1095 (2025). https://doi.org/10.1038/s41586-024-08470-1

24 Ju, L., MacDonald, A. H., Mak, K. F., Shan, J. & Xu, X. The fractional quantum anomalous Hall effect. *Nature Reviews Materials* **9**, 455-459 (2024). https://doi.org/10.1038/s41578-024-00694-x

25 Zhao, Y. *et al.* Automated processing and transfer of two-dimensional materials with robotics. *Nature Chemical Engineering* **2**, 296-308 (2025). https://doi.org/10.1038/s44286-025-00227-5

26 Liu, H., Zhao, J. & Ly, T. H. Clean Transfer of Two-Dimensional Materials: A Comprehensive Review. *ACS Nano* **18**, 11573-11597 (2024). https://doi.org/10.1021/acsnano.4c01000

27 Zhang, S. *et al.* Two-dimensional heterostructures and their device applications: progress, challenges and opportunities—review. *J. Phys. D: Appl. Phys* **54** (2021). https://doi.org/10.1088/1361-6463/ac16a4

28 Schranghamer, T. F., Sharma, M., Singh, R. & Das, S. Review and comparison of layer transfer methods for two-dimensional materials for emerging applications. *Chem. Soc. Rev.* **50**, 11032-11054 (2021). https://doi.org/10.1039/d1cs00706h

29 He, J. *et al.* 2D van der Waals heterostructures: processing, optical properties and applications in ultrafast photonics. *Mater. Horiz.* **7**, 2903-2921 (2020). https://doi.org/10.1039/d0mh00340a

30 Frisenda, R. *et al.* Recent progress in the assembly of nanodevices and van der Waals heterostructures by deterministic placement of 2D materials. *Chem. Soc. Rev.* **47**, 53-68 (2018). https://doi.org/10.1039/c7cs00556c

31 Cheliotis, I. & Zergioti, I. A review on transfer methods of two-dimensional materials. *2D Mater.* **11** (2024). https://doi.org/10.1088/2053-1583/ad2f43

32 Somphonsane, R., Buapan, K. & Ramamoorthy, H. Advances in 2D Material Transfer Systems for van der Waals Heterostructure Assembly. *Appl. Sci.* **14** (2024). https://doi.org/10.3390/app14146341

33 McKenzie, J., Sharma, N. & Liu, X. Fabrication of pristine 2D heterostructures for scanning probe microscopy. *APL Mater.* **12** (2024). https://doi.org/10.1063/5.0213542

34 Zomer, P. J., Guimarães, M. H. D., Brant, J. C., Tombros, N. & van Wees, B. J. Fast pick up technique for high quality heterostructures of bilayer graphene and hexagonal boron nitride. *Appl. Phys. Lett.* **105** (2014). https://doi.org/10.1063/1.4886096

35 Pizzocchero, F. *et al.* The hot pick-up technique for batch assembly of van der Waals heterostructures. *Nat. Commun.* **7**, 11894 (2016). https://doi.org/10.1038/ncomms11894

36 Kinoshita, K. *et al.* Dry release transfer of graphene and few-layer h-BN by utilizing thermoplasticity of polypropylene carbonate. *npj 2D Mater. and Appl* **3** (2019). https://doi.org/10.1038/s41699-019-0104-8

37 Kim, K. *et al.* van der Waals Heterostructures with High Accuracy Rotational Alignment. *Nano Lett.* **16**, 1989-1995 (2016). https://doi.org/10.1021/acs.nanolett.5b05263

38 Purdie, D. G. *et al.* Cleaning interfaces in layered materials heterostructures. *Nat. Commun.* **9**, 5387 (2018). https://doi.org/10.1038/s41467-018-07558-3

39 You, Y., Ni, Z., Yu, T. & Shen, Z. Edge chirality determination of graphene by Raman spectroscopy. *Appl. Phys. Lett.* **93** (2008). https://doi.org/10.1063/1.3005599

40 Xu, B., Mao, N., Zhao, Y., Tong, L. & Zhang, J. Polarized Raman Spectroscopy for Determining Crystallographic Orientation of Low-Dimensional Materials. *J. Phys. Chem. Lett.* **12**, 7442-7452 (2021). https://doi.org/10.1021/acs.jpclett.1c01889

41 Zobeiri, H., Wang, R., Deng, C., Zhang, Q. & Wang, X. Polarized Raman of Nanoscale Two-Dimensional Materials: Combined Optical and Structural Effects. *J. Phys. Chem. C* **123**, 23236-23245 (2019). https://doi.org/10.1021/acs.jpcc.9b06892

42 Huang, C. W. *et al.* Revealing anisotropic strain in exfoliated graphene by polarized Raman spectroscopy. *Nanoscale* **5**, 9626-9632 (2013). https://doi.org/10.1039/c3nr00123g

43 Kumar, N. *et al.* Second harmonic microscopy of monolayer MoS2. *Physical Review B* **87** (2013). https://doi.org/10.1103/PhysRevB.87.161403

44 Ribeiro-Soares, J. *et al.* Second Harmonic Generation in $WSe_2$. *2D Mater.* **2** (2015). https://doi.org/10.1088/2053-1583/2/4/045015

45 Yin, X. *et al.* Edge nonlinear optics on a $MoS_2$ atomic monolayer. *Science* **344**, 488-490 (2014). https://doi.org/10.1126/science.1250564

46 W, B. R. *Nonlinear Optics*. (Elsevier, 2008).

47 Debnath, R., Sett, S., Biswas, R., Raghunathan, V. & Ghosh, A. A simple fabrication strategy for orientationally accurate twisted heterostructures. *Nanotech.* **32** (2021). https://doi.org/10.1088/1361-6528/ac1756

48 Rosenberger, M. R. *et al.* Nano-"Squeegee" for the Creation of Clean 2D Material Interfaces. *ACS Appl. Mater. Interfaces* **10**, 10379-10387 (2018). https://doi.org/10.1021/acsami.8b01224

49 Jin, C. *et al.* Observation of moire excitons in $WSe_2/WS_2$ heterostructure superlattices. *Nature* **567**, 76-80 (2019). https://doi.org/10.1038/s41586-019-0976-y

50 Jessen, B. S. *et al.* Lithographic band structure engineering of graphene. *Nature Nanotech.* **14**, 340-346 (2019). https://doi.org/10.1038/s41565-019-0376-3

51 Lyons, T. P. *et al.* Interplay between spin proximity effect and charge-dependent exciton dynamics in $MoSe_2/CrBr_3$ van der Waals heterostructures. *Nat. Commun.* **11**, 6021 (2020). https://doi.org/10.1038/s41467-020-19816-4

52 Li, H. *et al.* Imaging two-dimensional generalized Wigner crystals. *Nature* **597**, 650-654 (2021). https://doi.org/10.1038/s41586-021-03874-9

53 Smolenski, T. *et al.* Signatures of Wigner crystal of electrons in a monolayer semiconductor. *Nature* **595**, 53-57 (2021). https://doi.org/10.1038/s41586-021-03590-4

54 Guo, Y. *et al.* Superconductivity in 5.0 degrees twisted bilayer $WSe_2$. *Nature* **637**, 839-845 (2025). https://doi.org/10.1038/s41586-024-08381-1

55 Cohen, L. A. *et al.* Nanoscale electrostatic control in ultraclean van der Waals heterostructures by local anodic oxidation of graphite gates. *Nat. Phys.* **19**, 1502-1508 (2023). https://doi.org/10.1038/s41567-023-02114-3

56 Jin, K. *et al.* Assembly of Arbitrary Designer Heterostructures with Atomically Clean Interfaces. *Adv. Mater. Interfaces* **11** (2023). https://doi.org/10.1002/admi.202300658

57 Wang, W. *et al.* Clean assembly of van der Waals heterostructures using silicon nitride membranes. *Nat. Electron* **6**, 981-990 (2023). https://doi.org/10.1038/s41928-023-01075-y

58 Shah, S. J. *et al.* Progress and prospects of Moiré superlattices in twisted TMD heterostructures. *Nano Res.* **17**, 10134-10161 (2024). https://doi.org/10.1007/s12274-024-6936-3

59 Shi, Y., Yamamoto, E., Kobayashi, M. & Osada, M. Automated One-Drop Assembly for Facile 2D Film Deposition. *ACS Appl. Mater. Interfaces* **15**, 22737-22743 (2023). https://doi.org/10.1021/acsami.3c02250

60 Masubuchi, S. *et al.* Autonomous robotic searching and assembly of two-dimensional crystals to build van der Waals superlattices. *Nat. Commun.* **9**, 1413 (2018). https://doi.org/10.1038/s41467-018-03723-w

61 Andrei, E. Y. *et al.* The marvels of moiré materials. *Nature Reviews Materials* **6**, 201-206 (2021). https://doi.org/10.1038/s41578-021-00284-1

62 Zhou, R. *et al.* Twisto-photonics in two-dimensional materials: A comprehensive review. *Nanotechnol. Rev.* **13** (2024). https://doi.org/10.1515/ntrev-2024-0086

63 Li, Y., Kuang, G., Jiao, Z., Yao, L. & Duan, R. Recent progress on the mechanical exfoliation of 2D transition metal dichalcogenides. *Materials Research Express* **9** (2022). https://doi.org/10.1088/2053-1591/aca6c6

64 Huang, Y. *et al.* Universal mechanical exfoliation of large-area 2D crystals. *Nat. Commun.* **11**, 2453 (2020). https://doi.org/10.1038/s41467-020-16266-w

65 Huang, Y. *et al.* Reliable Exfoliation of Large-Area High-Quality Flakes of Graphene and Other Two-Dimensional Materials. *ACS Nano* **9**, 10612-10620 (2015). https://doi.org/10.1021/acsnano.5b04258

66 DiCamillo, K., Krylyuk, S., Shi, W., Davydov, A. & Paranjape, M. Automated Mechanical Exfoliation of MoS2 and MoTe2 Layers for Two-Dimensional Materials Applications. *IEEE Transactions on Nanotechnology* **18**, 144-148 (2019). https://doi.org/10.1109/tnano.2018.2868672

67 Kobayashi, T., Sato, C., Dohi, T. & Kiriya, D. Propose an automated exfoliation process of $MoS_2$ with a universal mechanical setup. *Appl. Phys. Express* **16** (2023). https://doi.org/10.35848/1882-0786/acfd7d

68 Courtney, E. D. S., Pendharkar, M., Bittner, N. J., Sharpe, A. L. & Goldhaber-Gordon, D. Automated tabletop exfoliation and identification of monolayer graphene flakes. *Rev. Sci. Instrum.* **96** (2025). https://doi.org/10.1063/5.0255656

69 Gasbarro, A., Masuda, Y.-S. D., Ordonez, R. C., Weldon, J. A. & Lubecke, V. M. Accessible and Inexpensive Parameter Testing Platform for Adhesive Removal in Mechanical Exfoliation Procedures. *Electronics* **14** (2025). https://doi.org/10.3390/electronics14030533

70 Sozen, Y., Riquelme, J. J., Xie, Y., Munuera, C. & Castellanos-Gomez, A. High-Throughput Mechanical Exfoliation for Low-Cost Production of van der Waals Nanosheets. *Small Methods* **7**, e2300326 (2023). https://doi.org/10.1002/smtd.202300326

71 Shrestha, S. *et al.* Room temperature valley polarization via spin selective charge transfer. *Nat. Commun.* **14**, 5234 (2023). https://doi.org/10.1038/s41467-023-40967-7

72 Karmakar, A. *et al.* Excitation-Dependent High-Lying Excitonic Exchange via Interlayer Energy Transfer from Lower-to-Higher Bandgap 2D Material. *Nano Lett.* **23**, 5617-5624 (2023). https://doi.org/10.1021/acs.nanolett.3c01127

73 Huang, Z. *et al.* Observation of half-integer Shapiro steps in graphene Josephson junctions. *Appl. Phys. Lett.* **122** (2023). https://doi.org/10.1063/5.0153646

74 Huang, Z. *et al.* The study of contact properties in edge-contacted graphene–aluminum Josephson junctions. *Appl. Phys. Lett.* **121** (2022). https://doi.org/10.1063/5.0135034

75 Huang, Z. *et al.* Mechanisms of Interface Cleaning in Heterostructures Made from Polymer-Contaminated Graphene. *Small* **18**, e2201248 (2022). https://doi.org/10.1002/smll.202201248

76 Dai, Z. *et al.* Quantum-Well Bound States in Graphene Heterostructure Interfaces. *Phys. Rev. Lett.* **127**, 086805 (2021). https://doi.org/10.1103/PhysRevLett.127.086805

77 Shin, Y. J. *et al.* Fast and accurate robotic optical detection of exfoliated graphene and hexagonal boron nitride by deep neural networks. *2D Mater.* **8** (2021). https://doi.org/10.1088/2053-1583/abd72c

78 Gu, L. *et al.* Giant optical nonlinearity of Fermi polarons in atomically thin semiconductors. *Nat. Photon.* **18**, 816-822 (2024). https://doi.org/10.1038/s41566-024-01434-x

79 Tallon, B., Lipton-Duffin, J. & MacLeod, J. Exfoliation of Graphene onto Si(111)-7 x 7 under Ultrahigh Vacuum Provides Some Protection against Exposure to Air. *Langmuir* **40**, 25692-25697 (2024). https://doi.org/10.1021/acs.langmuir.4c03712

80 Haider, G. *et al.* Highly Efficient Bulk-Crystal-Sized Exfoliation of 2D Materials under Ultrahigh Vacuum. *ACS Appl. Electron. Mater.* **6**, 2301-2308 (2024). https://doi.org/10.1021/acsaelm.3c01824

81 Grubisic-Cabo, A. *et al.* In Situ Exfoliation Method of Large-Area 2D Materials. *Adv. Sci.* **10**, e2301243 (2023). https://doi.org/10.1002/advs.202301243

82 Sun, Z. *et al.* Exfoliation of 2D van der Waals crystals in ultrahigh vacuum for interface engineering. *Sci. Bull.* **67**, 1345-1351 (2022). https://doi.org/10.1016/j.scib.2022.05.017

83 Paton, K. R. *et al.* Scalable production of large quantities of defect-free few-layer graphene by shear exfoliation in liquids. *Nat. Mater.* **13**, 624-630 (2014). https://doi.org/10.1038/nmat3944

84 Dong, X. *et al.* Line-Scan Hyperspectral Imaging Microscopy with Linear Unmixing for Automated Two-Dimensional Crystals Identification. *ACS Photonics* **7**, 1216-1225 (2020). https://doi.org/10.1021/acsphotonics.0c00050

85 Dong, X. *et al.* 3D Deep Learning Enables Accurate Layer Mapping of 2D Materials. *ACS Nano* **15**, 3139-3151 (2021). https://doi.org/10.1021/acsnano.0c09685

86 Chang, Y. C., Wang, Y. K., Chen, Y. T. & Lin, D. Y. Facile and Reliable Thickness Identification of Atomically Thin Dichalcogenide Semiconductors Using Hyperspectral Microscopy. *Nanomaterials* **10** (2020). https://doi.org/10.3390/nano10030526

87 Li, K. C. *et al.* Intelligent Identification of $MoS_2$ Nanostructures with Hyperspectral Imaging by 3D-CNN. *Nanomaterials* **10** (2020). https://doi.org/10.3390/nano10061161

88 Funke, S. *et al.* Spectroscopic imaging ellipsometry for automated search of flakes of mono- and n-layers of 2D-materials. *Appl. Surf. Sci.* **421**, 435-439 (2017). https://doi.org/10.1016/j.apsusc.2016.10.158

89 Uslu, J.-L. *et al.* An open-source robust machine learning platform for real-time detection and classification of 2D material flakes. *Mach. Learn.: Sci. Technol.* **5** (2024). https://doi.org/10.1088/2632-2153/ad2287

90 Saito, Y. *et al.* Deep-learning-based quality filtering of mechanically exfoliated 2D crystals. *npj Comput. Mater.* **5** (2019). https://doi.org/10.1038/s41524-019-0262-4

91 Masubuchi, S. *et al.* Deep-learning-based image segmentation integrated with optical microscopy for automatically searching for two-dimensional materials. *npj 2D Mater. and Appl* **4** (2020). https://doi.org/10.1038/s41699-020-0137-z

92 Ramezani, F. *et al.* Automatic detection of multilayer hexagonal boron nitride in optical images using deep learning-based computer vision. *Sci. Rep.* **13**, 1595 (2023). https://doi.org/10.1038/s41598-023-28664-3

93 Zenebe, Y. A. *et al.* in *2022 19th International Computer Conference on Wavelet Active Media Technology and Information Processing (ICCWAMTIP).* (IEEE Xplore).

94 Siao, H.-Y. *et al.* Machine Learning-based Automatic Graphene Detection with Color Correction for Optical Microscope Images. (2021). <https://arxiv.org/abs/2103.13495>.

95 Sanchez-Juarez, J., Granados-Baez, M., Aguilar-Lasserre, A. A. & Cardenas, J. Automated system for the detection of 2D materials using digital image processing and deep learning. *Opt. Mater. Express* **12** (2022). https://doi.org/10.1364/ome.454314

96 Zhang, Y., Zhang, H., Zhou, S., Liu, G. & Zhu, J. Deep Learning-Based Layer Identification of 2D Nanomaterials. *Coatings* **12** (2022). https://doi.org/10.3390/coatings12101551

97 Wu, B., Wang, L. & Gao, Z. in *2019 International Conference on Information Technology and Computer Application (ITCA).* 247-252.

98 Leger, P. A., Ramesh, A., Ulloa, T. & Wu, Y. Machine learning enabled fast optical identification and characterization of 2D materials. *Sci. Rep.* **14**, 27808 (2024). https://doi.org/10.1038/s41598-024-79386-z

99 Sterbentz, R. M., Haley, K. L. & Island, J. O. Universal image segmentation for optical identification of 2D materials. *Sci. Rep.* **11**, 5808 (2021). https://doi.org/10.1038/s41598-021-85159-9

100 Yang, E. *et al.* Machine Learning-Assisted Identification of Single-Layer Graphene via Color Variation Analysis. *Nanomaterials* **14** (2024). https://doi.org/10.3390/nano14020183

101 Cho, W. H., Shin, J., Kim, Y. D. & Jung, G. J. Pixel-wise classification in graphene-detection with tree-based machine learning algorithms. *Mach. Learn.: Sci. Technol.* **3** (2022). https://doi.org/10.1088/2632-2153/aca744

102 Nguyen, X. B. *et al.* Two-Dimensional Quantum Material Identification via Self-Attention and Soft-Labeling in Deep Learning. *IEEE Access* **12**, 139683-139691 (2024). https://doi.org/10.1109/access.2024.3465221

103 Mahjoubi, S., Ye, F., Bao, Y., Meng, W. & Zhang, X. Identification and classification of exfoliated graphene flakes from microscopy images using a hierarchical deep convolutional neural network. *Eng. Appl. Artif. Intell.* **119** (2023). https://doi.org/10.1016/j.engappai.2022.105743

104 Lin, X. *et al.* Intelligent identification of two-dimensional nanostructures by machine-learning optical microscopy. *Nano Res.* **11**, 6316-6324 (2018). https://doi.org/10.1007/s12274-018-2155-0

105 Bhawsar, S., Fang, M., Sarkar, A. S., Chen, S. & Yang, E.-H. Deep learning-based multimodal analysis for transition-metal dichalcogenides. *MRS Bulletin* **49**, 1021-1031 (2024). https://doi.org/10.1557/s43577-024-00741-6

106 Masubuchi, S. & Machida, T. Classifying optical microscope images of exfoliated graphene flakes by data-driven machine learning. *npj 2D Mater. and Appl* **3** (2019). https://doi.org/10.1038/s41699-018-0084-0

107 Zichi, L., Liu, T., Drueke, E., Zhao, L. & Xu, G. Physically informed machine-learning algorithms for the identification of two-dimensional atomic crystals. *Sci. Rep.* **13**, 6143 (2023). https://doi.org/10.1038/s41598-023-33298-6

108 Greplova, E. *et al.* Fully Automated Identification of Two-Dimensional Material Samples. *Phys. Rev. Appl* **13** (2020). https://doi.org/10.1103/PhysRevApplied.13.064017

109 Dong, X. *et al.* Deep-Learning-Based Microscopic Imagery Classification, Segmentation, and Detection for the Identification of 2D Semiconductors. *Advanced Theory and Simulations* **5** (2022). https://doi.org/10.1002/adts.202200140

110 Han, B. *et al.* Deep-Learning-Enabled Fast Optical Identification and Characterization of 2D Materials. *Adv. Mater.* **32**, e2000953 (2020). https://doi.org/10.1002/adma.202000953

111 Mondal, M., Dash, A. K. & Singh, A. Optical Microscope Based Universal Parameter for Identifying Layer Number in Two-Dimensional Materials. *ACS Nano* **16**, 14456-14462 (2022). https://doi.org/10.1021/acsnano.2c04833

112 Lee, J. *et al.* Highly efficient computer algorithm for identifying layer thickness of atomically thin 2D materials. *J. Phys. D: Appl. Phys* **51** (2018). https://doi.org/10.1088/1361-6463/aaac19

113 Sahriar, M. A. *et al.* Versatile recognition of graphene layers from optical images under controlled illumination through green channel correlation method. *Nanotech.* **34** (2023). https://doi.org/10.1088/1361-6528/ace979

114 Hu, X., Qiu, C. & Liu, D. Rapid thin-layer $WS_2$ detection based on monochromatic illumination photographs. *Nano Res.* **14**, 840-845 (2020). https://doi.org/10.1007/s12274-020-3122-0

115 Balasubramaniyan, S. *et al.* in *Computational Collective Intelligence: 13th International Conference, ICCCI 2021, Rhodes, Greece, September 29 – October 1, 2021, Proceedings* 784–791 (Springer-Verlag, Rhodos, Greece, 2021).

116 Anzai, Y. *et al.* Broad range thickness identification of hexagonal boron nitride by colors. *Appl. Phys. Express* **12** (2019). https://doi.org/10.7567/1882-0786/ab0e45

117 Taghavi, N. S. *et al.* Thickness determination of $MoS_2$, $MoSe_2$, $WS_2$ and $WSe_2$ on transparent stamps used for deterministic transfer of 2D materials. *Nano Res.* **12**, 1691-1695 (2019). https://doi.org/10.1007/s12274-019-2424-6

118 Joy, N. J., K, R. M. & Balakrishnan, J. A simple and robust machine learning assisted process flow for the layer number identification of TMDs using optical contrast spectroscopy. *J. Phys. Condens. Matter* **51** (2022). https://doi.org/10.1088/1361-648X/ac9f96

119 *Brookhaven National Laboratory QPress*, <https://www.bnl.gov/qpress/> (2025).

120 Hattori, Y., Taniguchi, T., Watanabe, K. & Kitamura, M. Identification of exfoliated monolayer hexagonal boron nitride films with a digital color camera under white light illumination. *Nanotech.* **35** (2024). https://doi.org/10.1088/1361-6528/ad58e7

121 Hattori, Y., Taniguchi, T., Watanabe, K. & Kitamura, M. Antireflection Substrates for Determining the Number of Layers of Few-Layer Hexagonal Boron Nitride Films and for Visualizing Organic Monolayers. *ACS Appl. Nano Mater.* **6**, 21876-21886 (2023). https://doi.org/10.1021/acsanm.3c04075

122 Hattori, Y., Taniguchi, T., Watanabe, K. & Kitamura, M. Visualization of a hexagonal born nitride monolayer on an ultra-thin gold film via reflected light microscopy. *Nanotech.* **33** (2021). https://doi.org/10.1088/1361-6528/ac3357

123 Chandrasekar, H. *et al.* Spotting 2D atomic layers on aluminum nitride thin films. *Nanotech.* **26**, 425202 (2015). https://doi.org/10.1088/0957-4484/26/42/425202

124 Hutzler, A. *et al.* Large-Area Layer Counting of Two-Dimensional Materials Evaluating the Wavelength Shift in Visible-Reflectance Spectroscopy. *J. Phys. Chem. C* **123**, 9192-9201 (2019). https://doi.org/10.1021/acs.jpcc.9b00957

125 Hutzler, A., Fritsch, B., Matthus, C. D., Jank, M. P. M. & Rommel, M. Highly accurate determination of heterogeneously stacked Van-der-Waals materials by optical microspectroscopy. *Sci. Rep.* **10**, 13676 (2020). https://doi.org/10.1038/s41598-020-70580-3

126 Nguyen, D. C. *et al.* Visibility of hexagonal boron nitride on transparent substrates. *Nanotech.* **31**, 195701 (2020). https://doi.org/10.1088/1361-6528/ab6bf4

127 Romagnoli, P. *et al.* Making graphene visible on transparent dielectric substrates: Brewster angle imaging. *2D Mater.* **2** (2015). https://doi.org/10.1088/2053-1583/2/3/035017

128 Wang, Z., Lin, Y. C., Zhang, K., Wu, W. & Huang, S. Measuring complex refractive index through deep-learning-enabled optical reflectometry. *2D Mater.* **10** (2023). https://doi.org/10.1088/2053-1583/acc59b

129 Simsek, E. Determining optical constants of 2D materials with neural networks from multi-angle reflectometry data. *Mach. Learn.: Sci. Technol.* **1** (2020). https://doi.org/10.1088/2632-2153/ab6d5f

130 Yang, J. & Yao, H. Automated identification and characterization of two-dimensional materials via machine learning-based processing of optical microscope images. *Extreme Mechanics Letters* **39** (2020). https://doi.org/10.1016/j.eml.2020.100771

131 Mao, Y. *et al.* Identification of triangular single crystals of transition metal dichalcogenides based on the detection algorithm. *Opt. Lett.* **49**, 298-301 (2024). https://doi.org/10.1364/OL.510325

132 Yang, H. *et al.* Identification and Structural Characterization of Twisted Atomically Thin Bilayer Materials by Deep Learning. *Nano Lett.* **24**, 2789-2797 (2024). https://doi.org/10.1021/acs.nanolett.3c04815

133 Li, Z., Lee, J., Yao, F. & Sun, H. Quantifying the CVD-grown two-dimensional materials via image clustering. *Nanoscale* **13**, 15324-15333 (2021). https://doi.org/10.1039/d1nr03802h

134 Solís-Fernández, P. & Ago, H. Machine Learning Determination of the Twist Angle of Bilayer Graphene by Raman Spectroscopy: Implications for van der Waals Heterostructures. *ACS Appl. Nano Mater.* **5**, 1356-1366 (2022). https://doi.org/10.1021/acsanm.1c03928

135 *GitHub - Jaluus/2DMatGMM*, <https://github.com/Jaluus/2DMatGMM> (2025).

136 *GitHub - uark-cviu/quantumflake*, <https://github.com/uark-cviu/quantumflake > (2025).

137 Pham, P. V. *et al.* Transfer of 2D Films: From Imperfection to Perfection. *ACS Nano* **18**, 14841-14876 (2024). https://doi.org/10.1021/acsnano.4c00590

138 Fan, S., Li, X., Mondal, A., Wang, W. & Lee, Y. H. Strategy for transferring van der Waals materials and heterostructures. *2D Mater.* **11** (2024). https://doi.org/10.1088/2053-1583/ad4044

139 Fox, C., Mao, Y., Zhang, X., Wang, Y. & Xiao, J. Stacking Order Engineering of Two-Dimensional Materials and Device Applications. *Chem. Rev.* **124**, 1862-1898 (2024). https://doi.org/10.1021/acs.chemrev.3c00618

140 Craig, I. M. *et al.* Local atomic stacking and symmetry in twisted graphene trilayers. *Nat. Mater.* **23**, 323-330 (2024). https://doi.org/10.1038/s41563-023-01783-y

141 Barré, E. *et al.* Engineering interlayer hybridization in van der Waals bilayers. *Nature Reviews Materials* **9**, 499-508 (2024). https://doi.org/10.1038/s41578-024-00666-1

142 Wang, Q., Wang, X., Lou, Q., Jiang, Y. & Fan, X. Two-Dimensional Spiral: A Promising Moiré Superlattice. *Laser & Photonics Reviews* **19** (2024). https://doi.org/10.1002/lpor.202401368

143 Han, Z., Wang, F., Sun, J., Wang, X. & Tang, Z. Recent Advances in Ultrathin Chiral Metasurfaces by Twisted Stacking. *Adv. Mater.* **35**, e2206141 (2023). https://doi.org/10.1002/adma.202206141

144 Lau, C. N., Bockrath, M. W., Mak, K. F. & Zhang, F. Reproducibility in the fabrication and physics of moire materials. *Nature* **602**, 41-50 (2022). https://doi.org/10.1038/s41586-021-04173-z

145 Hong, K. S., Chen, O. & Bai, Y. Emergent quantum properties from low-dimensional building blocks and their superlattices. *Nano Res.* **17**, 10490-10510 (2024). https://doi.org/10.1007/s12274-024-6984-8

146 Mannix, A. J. *et al.* Robotic four-dimensional pixel assembly of van der Waals solids. *Nature Nanotech.* **17**, 361-366 (2022). https://doi.org/10.1038/s41565-021-01061-5

147 Ahn, S.-H. *et al.* in *Volume 2: Manufacturing Equipment and Automation; Manufacturing Processes; Manufacturing Systems; Nano/Micro/Meso Manufacturing; Quality and Reliability* (American Society of Mechanical Engineers, 2023).

148 Masubuchi, S. *et al.* Dry pick-and-flip assembly of van der Waals heterostructures for microfocus angle-resolved photoemission spectroscopy. *Sci. Rep.* **12**, 10936 (2022). https://doi.org/10.1038/s41598-022-14845-z

149 Moon, D. *et al.* Hypotaxy of wafer-scale single-crystal transition metal dichalcogenides. *Nature* **638**, 957-964 (2025). https://doi.org/10.1038/s41586-024-08492-9

150 Chen, L. *et al.* Fully automatic transfer and measurement system for structural superlubric materials. *Nat. Commun.* **14**, 6323 (2023). https://doi.org/10.1038/s41467-023-41859-6

151 Han, S. S. *et al.* Automated Assembly of Wafer-Scale 2D TMD Heterostructures of Arbitrary Layer Orientation and Stacking Sequence Using Water Dissoluble Salt Substrates. *Nano Lett.* **20**, 3925-3934 (2020). https://doi.org/10.1021/acs.nanolett.0c01089

152 Meng, Y. *et al.* Functionalizing nanophotonic structures with 2D van der Waals materials. *Nanoscale Horiz.* **8**, 1345-1365 (2023). https://doi.org/10.1039/d3nh00246b

153 Corbett, B., Loi, R., Zhou, W., Liu, D. & Ma, Z. Transfer print techniques for heterogeneous integration of photonic components. *Progress in Quantum Electronics* **52**, 1-17 (2017). https://doi.org/10.1016/j.pquantelec.2017.01.001

154 You, J. *et al.* Hybrid/Integrated Silicon Photonics Based on 2D Materials in Optical Communication Nanosystems. *Laser & Photonics Reviews* **14** (2020). https://doi.org/10.1002/lpor.202000239

155 Hu, Z., Sun, R.-X., Chen, X.-D., Tian, J. & Liu, Z. Pattern Transfer for van der Waals Integration. *ACS Appl. Electron. Mater.* **6**, 8463-8473 (2024). https://doi.org/10.1021/acsaelm.4c01664

156 Nguyen, V. *et al.* Deterministic Assembly of Arrays of Lithographically Defined $WS_2$ and $MoS_2$ Monolayer Features Directly From Multilayer Sources Into Van Der Waals Heterostructures. *Journal of Micro and Nano-Manufacturing* **7** (2019). https://doi.org/10.1115/1.4045259

157 Long, G. *et al.* Achieving Ultrahigh Carrier Mobility in Two-Dimensional Hole Gas of Black Phosphorus. *Nano Lett.* **16**, 7768-7773 (2016). https://doi.org/10.1021/acs.nanolett.6b03951

158 Jimenez-Arevalo, N. *et al.* X-ray photoelectron spectroscopy of high-throughput mechanically exfoliated van der Waals materials. *Nanoscale* **16**, 17559-17566 (2024). https://doi.org/10.1039/d4nr02882a

159 Patil, V. *et al.* Pick-up and assembling of chemically sensitive van der Waals heterostructures using dry cryogenic exfoliation. *Sci. Rep.* **14**, 11097 (2024). https://doi.org/10.1038/s41598-024-58935-6

160 Mukherjee, S. *et al.* Toward Phonon-Limited Transport in Two-Dimensional Transition Metal Dichalcogenides by Oxygen-Free Fabrication. *ACS Nano* **19**, 9327-9339 (2025). https://doi.org/10.1021/acsnano.5c00995

161 Gao, J. *et al.* Aging of Transition Metal Dichalcogenide Monolayers. *ACS Nano* **10**, 2628-2635 (2016). https://doi.org/10.1021/acsnano.5b07677

162 Park, J. H. *et al.* Selective Chemical Response of Transition Metal Dichalcogenides and Metal Dichalcogenides in Ambient Conditions. *ACS Appl. Mater. Interfaces* **9**, 29255-29264 (2017). https://doi.org/10.1021/acsami.7b08244

163 Nguyen, V. H. *et al.* Fast fabrication technique for high-quality van der Waals heterostructures using inert shielding gas environment. *Appl. Surf. Sci.* **639** (2023). https://doi.org/10.1016/j.apsusc.2023.158186

164 Gray, M. J. *et al.* A cleanroom in a glovebox. *Rev. Sci. Instrum.* **91**, 073909 (2020). https://doi.org/10.1063/5.0006462

165 Duleba, A. *et al.* Inert-Atmosphere Microfabrication Technology for 2D Materials and Heterostructures. *Micromachines* **15** (2023). https://doi.org/10.3390/mi15010094

166 Gant, P. *et al.* A system for the deterministic transfer of 2D materials under inert environmental conditions. *2D Mater.* **7** (2020). https://doi.org/10.1088/2053-1583/ab72d6

167 Thompson, J. P., Doha, M. H., Murphy, P., Hu, J. & Churchill, H. O. H. Exfoliation and Analysis of Large-area, Air-Sensitive Two-Dimensional Materials. *JoVE* (2019). https://doi.org/10.3791/58693

168 Onodera, M. & Machida, T. Transfer of van der Waals Heterostructures of Two-dimensional Materials onto Microelectromechanical Systems. *Sensors and Materials* **35** (2023). https://doi.org/10.18494/sam4363

169 Kim, D. *et al.* Full-dry flipping transfer method for van der waals heterostructure. *Current Applied Physics* **59**, 165-168 (2024). https://doi.org/10.1016/j.cap.2023.10.018

170 Pasztor, A., Scarfato, A. & Renner, C. Note: Mechanical in situ exfoliation of van der Waals materials. *Rev. Sci. Instrum.* **88**, 076104 (2017). https://doi.org/10.1063/1.4993738

171 Guo, S. *et al.* An ultra-high vacuum system for fabricating clean two-dimensional material devices. *Rev. Sci. Instrum.* **94**, 013903 (2023). https://doi.org/10.1063/5.0110875

172 Merk, D., Rusponi, S. & Brune, H. Large Area Monolayer Graphene Transfer in Ultra-High Vacuum. *J. Phys. Chem. C* **129**, 7868-7878 (2025). https://doi.org/10.1021/acs.jpcc.4c08196

173 Imamura, H. *et al.* Twisted bilayer graphene fabricated by direct bonding in a high vacuum. *Appl. Phys. Express* **13** (2020). https://doi.org/10.35848/1882-0786/ab99d1

174 El-Kerdi, B. *et al.* Evidence of Strong Dzyaloshinskii-Moriya Interaction at the Cobalt/Hexagonal Boron Nitride Interface. *Nano Lett.* **23**, 3202-3208 (2023). https://doi.org/10.1021/acs.nanolett.2c04985

175 Liu, F. Mechanical exfoliation of large area 2D materials from vdW crystals. *Progress in Surface Science* **96** (2021). https://doi.org/10.1016/j.progsurf.2021.100626

176 Heyl, M. & List-Kratochvil, E. J. W. Only gold can pull this off: mechanical exfoliations of transition metal dichalcogenides beyond scotch tape. *Applied Physics A* **129** (2022). https://doi.org/10.1007/s00339-022-06297-z

177 Panasci, S. E., Schilirò, E., Roccaforte, F. & Giannazzo, F. Gold-Assisted Exfoliation of Large-Area Monolayer Transition Metal Dichalcogenides: From Interface Properties to Device Applications. *Advanced Functional Materials* **35** (2024). https://doi.org/10.1002/adfm.202414532

178 Pirker, L., Honolka, J., Velický, M. & Frank, O. When 2D materials meet metals. *2D Mater.* **11** (2024). https://doi.org/10.1088/2053-1583/ad286b

179 Kim, J. *et al.* Layer-resolved graphene transfer via engineered strain layers. *Science* **342**, 833-836 (2013). https://doi.org/10.1126/science.1242988

180 Desai, S. B. *et al.* Gold-Mediated Exfoliation of Ultralarge Optoelectronically-Perfect Monolayers. *Adv. Mater.* **28**, 4053-4058 (2016). https://doi.org/10.1002/adma.201506171

181 Heyl, M. *et al.* Low Temperature Heating of Silver-Mediated Exfoliation of MoS2. *Adv. Mater. Interfaces* **9** (2022). https://doi.org/10.1002/admi.202200362

182 Johnston, A. C. & Khondaker, S. I. Can Metals Other than Au be Used for Large Area Exfoliation of MoS2 Monolayers? *Adv. Mater. Interfaces* **9** (2022). https://doi.org/10.1002/admi.202200106

183 Lin, Z. *et al.* Controllable Growth of Large-Size Crystalline MoS2 and Resist-Free Transfer Assisted with a Cu Thin Film. *Sci Rep* **5**, 18596 (2015). https://doi.org/10.1038/srep18596

184 Shen, J. *et al.* Metal-assisted vacuum transfer enabling in situ visualization of charge density waves in monolayer MoS(2). *Sci Adv* **11**, eadr9753 (2025). https://doi.org/10.1126/sciadv.adr9753

185 Liu, F. *et al.* Disassembling 2D van der Waals crystals into macroscopic monolayers and reassembling into artificial lattices. *Science* **367**, 903-906 (2020). https://doi.org/10.1126/science.aba1416

186 Velicky, M. *et al.* Mechanism of Gold-Assisted Exfoliation of Centimeter-Sized Transition-Metal Dichalcogenide Monolayers. *ACS Nano* **12**, 10463-10472 (2018). https://doi.org/10.1021/acsnano.8b06101

187 Magda, G. Z. *et al.* Exfoliation of large-area transition metal chalcogenide single layers. *Sci Rep* **5**, 14714 (2015). https://doi.org/10.1038/srep14714

188 Galafassi, R. *et al.* Exfoliation and transfer of millimetre-sized MoS(2) flakes on arbitrary substrates. *Nanoscale Adv* (2025). https://doi.org/10.1039/d5na00919g

189 Zaborski, G., Jr. *et al.* Macroscopic Uniform 2D Moire Superlattices with Controllable Angles. *J. Am. Chem. Soc.* **147**, 38033-38042 (2025). https://doi.org/10.1021/jacs.5c09131

190 Olsen, N. *et al.* Macroscopic Transition Metal Dichalcogenide Monolayers from Gold-Tape Exfoliation Retain Intrinsic Properties. *Nano Lett* **25**, 15198-15205 (2025). https://doi.org/10.1021/acs.nanolett.5c03371

191 Kim, H. *et al.* Imaging inter-valley coherent order in magic-angle twisted trilayer graphene. *Nature* **623**, 942-948 (2023). https://doi.org/10.1038/s41586-023-06663-8

192 Yan, M., Wang, H., Cao, G. & Ren, W. Oxygen plasma activated Au-assisted mechanical exfoliation for large-area, high-quality metal phosphorus trichalcogenides flakes. *Chem. Phys. Lett.* **856** (2024). https://doi.org/10.1016/j.cplett.2024.141634

193 Tom, G. *et al.* Self-Driving Laboratories for Chemistry and Materials Science. *Chem. Rev.* **124**, 9633-9732 (2024). https://doi.org/10.1021/acs.chemrev.4c00055

194 He, Y., Yu, W. & Ouyang, G. Effect of stepped substrates on the interfacial adhesion properties of graphene membranes. *Phys. Chem. Chem. Phys.* **16**, 11390-11397 (2014). https://doi.org/10.1039/c4cp00633j

195 He, Y., Chen, W. F., Yu, W. B., Ouyang, G. & Yang, G. W. Anomalous interface adhesion of graphene membranes. *Sci. Rep.* **3**, 2660 (2013). https://doi.org/10.1038/srep02660